# Dynamics of type V menthol-thymol deep eutectic solvents: Do they reveal non-ideality?

Claire D'Hondt, Denis Morineau *

[†]Institute of Physics of Rennes, CNRS-University of Rennes 1, UMR 6251, F-35042 Rennes, France

ABSTRACT: We have established a comprehensive study of the molecular dynamics of the menthol-thymol mixture, the prototype of the new class of type V deep eutectic solvents. Dielectric spectroscopy and differential scanning calorimetry were combined to assess the dipolar relaxation and glassy dynamics over an extended range of timescales ($10^{-6}$ – $10^2$ s) and for eleven different compositions including the two pure constituents. Positive deviations from ideal mixing approximation were demonstrated for both the solvent dielectric strength and the glass transition. They support the idea that preferential H-bond interactions between unlike molecules induce structural rearrangements in the DES mixtures. Excellent glassforming capability of the DES mixtures was demonstrated for a broad range of compositions ($x_{\text{Thymol}}$ = 0.4-0.7). In this case, the dipolar dynamics could be assessed from the normal liquid to the deeply supercooled state. Many salient parameters of the relaxation functions, including fragility, non-Debye character and rotation-translation decoupling point to the development of dynamic heterogeneities.

**Corresponding Author:** * E-mail: denis.morineau@univ-rennes1.fr

# 1. Introduction

Deep eutectic solvents (DESs) have attracted growing interest as promising alternative to classical solvents.[1-6] The formation of DESs is usually achieved by mixing H-bonding molecules and molecular ionic species.[6] For eutectic mixtures, the melting point is notably depressed with respect to the pure ingredients.[7, 8] In DESs, this phenomenon can be reinforced by non-ideal mixing effects that result from preferential intermolecular interactions between unlike molecules in the mixtures.[9-11] This beneficial effect paves the way for designing solvents with specific properties by combining different ingredients which would otherwise be solid at room temperature.

While most common DESs incorporate at least one ionic ingredient, the recently introduced type V DESs are exclusively formed by non-ionic components.[12] Consequently, type V DESs are non-miscible with water, which contrasts with the former types of DESs.[13-15] As such, they are contemplated for extraction applications that require the use of hydrophobic solvents.[16-19]

Since 2019 and the seminal work of Abranches *et al.* [12], the non-ideal character of type V eutectic mixtures has been mostly investigated for terpene-based solvents.[20-25] For the prototypical menthol-thymol mixture, a large negative deviation from ideality was demonstrated from the experimental determination of the solid–liquid equilibrium phase diagram, allowing its classification as deep. The origin of non-ideality was attributed to the asymmetric character of H-bond interactions involving thymol, which behaves as a strong hydrogen bond donor (HBD) but a weak hydrogen bond acceptor (HBA).[12, 25] This condition favors the formation of H-bonds between thymol and menthol molecules, compared to the self-interactions (i.e. thymol-thymol and menthol-menthol H-bonds). This interpretation was supported by Raman spectroscopy and molecular dynamic simulation studies that underlined the fundamental role of hetero-association in determining the liquid structure of menthol-thymol mixtures.[24, 26, 27]

Interestingly, it was shown that the non-ideal character of the mixture was temperature dependent. The activity coefficient of thymol in the equimolar mixture varies considerably from less than 0.3 at 223 K to approximately 0.9 at 463 K.[24] Concomitantly, the number of H-bonds between the two DES-forming molecules decreases on increasing temperature.[24, 26] Besides temperature, varying the composition of the solvent is of primary interest. A maximum in the number of menthol-thymol H-bonds was found around the equimolar composition.[24] It is important to stress here that, although often misinterpreted in the literature, this does not imply that the eutectic composition can be associated with the existence of stoichiometric complexes (see detailed discussion by Martins *et al*. [8]). Rather, this observation suggests the existence of a range of compositions where the physico-chemical properties of the liquid mixture are affected by the preferential interactions between dissimilar molecules, and by the likely formation of mesoscopic liquid structures or dynamic heterogeneities. This advocates for more systematic studies of DES mixtures at compositions that are not restricted to the eutectic point or to specific stoichiometries. Incidentally, while the eutectic point of menthol-thymol mixtures was initially located close to the equimolar composition, a recent extensive study of their solid−liquid equilibria revealed a more complex situation. This updated phase diagram ruled out the standard single-eutectic point picture and demonstrated the existence of two co-crystals with stoichiometric ratios 1:3 and 3:2 and two eutectic points at respectively $x^{e1}_{\text{Thymol}}$ = 0.33 and $x^{e2}_{\text{Thymol}}$ = 0.48. [23]

So far, the fundamental understanding of type V menthol-thymol mixtures has been solely based on static (thermodynamic and structural) properties. The emerging questions that have not been addressed yet, concern the dynamics of type V DES. How does non-ideality, which appears intrinsically linked to the considerable decrease of the melting point and to the formation of specific H-bonded structures shows up in terms of molecular motions? More precisely, can one relate the observed important dependences of non-ideality on the temperature and on the composition to characteristic features in the relaxation processes of the liquid state? The aim of the present study is to provide an unprecedented assessment of the dynamical properties of a type V DES, covering an extended range of temperatures and compositions in the liquid, supercooled liquid, glassy and solid states. To reach this goal, dielectric spectroscopy (DS) is a perfectly well-suited method as it allows covering a broad range of frequencies (MHz–Hz) corresponding to the DES molecular dynamics, as already demonstrated in recent studied of ionic DESs based on choline chloride derivatives.[14, 28, 29] We combined DS to differential scanning calorimetry (DSC) in order to bridge the gap from short ($\sim 10^{-6}$ s) to ultra-long ($\sim 10^2$ s) timescales, that are typical for glassforming liquids.

## 2. Materials and methods

L-menthol (>99%) and thymol (>99%) were purchased from Sigma-Aldrich. DES and pure systems were prepared by weighting and adding the two precursors to produce a series of eleven samples with evenly spaced molar ratios in 0.1 steps from $x_{Thymol} = 0$ to $x_{Thymol} = 1$. The samples were mixed by mechanical agitation at about 60°C for 30 min until a clear homogeneous liquid phase was obtained.

For DS experiments, the sample was prepared in parallel plate geometry between two stainless steel electrodes with a diameter of 20 mm and a spacing of 260 μm maintained by Teflon spacers.

It was placed in the cryostat and maintained under a dry nitrogen atmosphere. The complex impedance of the as-prepared capacitor was measured from 1 Hz to $10^6$ Hz with a Novocontrol high resolution dielectric Alpha analyzer with an active sample cell. The measurements were performed at thermal equilibrium along a cooling and a subsequent heating branch with a temperature step of 2 K, and typically covering the temperature range from 130 K to 333 K (-143 °C to 60 °C). The temperature of the sample was controlled by a Quatro temperature controller (Novocontrol) with nitrogen as a heating/cooling agent providing a temperature stability within 0.1 K. The temperature scan rate, although discontinuous, was about 0.3 K.min$^{-1}$ on average.

For DSC experiments, the weighted samples were sealed in Tzero© aluminum hermetic pans. The measurements were performed with a Q-20 TA instrument equipped with a liquid nitrogen cooling system. The melting transition of an indium sample was used for calibration of temperature and heat flux. In general, the temperature was ramped linearly both on cooling and heating in the temperature range from 180 to 300 K with the same scanning rate of 5 K.min$^{-1}$. For concentrated systems ($x_{Thymol} \leq 0.1$ and $x_{Thymol} \geq 0.9$) the maximum temperature was increased to assure complete melting (up to 323-333 K) and the cooling rate was set to maximum speed (ca. 200 K.min$^{-1}$).

## 3. Results and discussion

*3.1 Differential scanning calorimetry*

The phase diagram of thymol-menthol DES has been studied in great details recently.[23] Due to polymorphism, metastability, and the presence of solid solutions, the determination of the phase diagram required the combination of X-ray diffraction and DSC, along with carefully designed thermal treatments. It demonstrated the existence of two co-crystals with stoichiometric ratios 1:3 and 3:2 and two regions of solid solutions formed by menthol in thymol and menthol in the 1:3 co-crystal. As a result, two eutectic points were established for compositions $x_{\text{Thymol}}^{E1} = 0.33$ and $x_{\text{Thymol}}^{E2} = 0.48$, respectively.[23] The present study focuses on the dynamics of the liquid states and it was not conceived to bring any new knowledge about the solid-liquid equilibria and the nature of the crystalline states. In fact, on the contrary, our intention is to avoid crystallization as often as possible. The aim of DSC experiments are twofold: - determine for each composition and thermal treatment, the temperature range where the solvent has remained entirely liquid (or glassy), – determine the calorimetric glass transition of the fully vitrified samples.

The thermograms measured for the eleven different compositions are illustrated in Figs. S1. The thermograms during heating (red curves) were all acquired at 5 K.min$^{-1}$ after an initial cooling branch (blue curves) with the same scanning rate, except for $x_{\text{Thymol}}$ = 1, 0.9, 0.1, and 0 (Figs. S1a, b, and Figs. S1j,k), where a fast cooling rate of 200 K.min$^{-1}$ was applied. Different thermal behaviors were observed, depending on the composition. Importantly, the formation of a glass of the entire solvent, and the subsequent determination of its glass transition $T_g$ on heating could be achieved for eight samples with $x_{\text{Thymol}} \geq 0.3$. This is demonstrated by the absence of exothermic peak during the entire cooling scan and also during subsequent heating up to $T_g$, which can be identified by the characteristic jump of the heat capacity. Only the three menthol-rich samples ($x_{\text{Thymol}} \leq 0.2$) are not glassforming systems under the applied experimental conditions. Indeed, crystallization occurred on cooling, as indicated by exothermic events in Figs. S1i-k even though fast cooling rates were applied.

As a whole, the mixtures can be classified in four different groups according to their thermal behavior observed by DSC:

- very good glassformers: for close to equimolar compositions, $0.4 \leq x_{Thymol} \leq 0.6$, crystallization never occurred, as illustrated in Figs. S1e-g.

- good glassformers: for intermediate compositions, $x_{Thymol} = 0.3$ (Fig. S1h) and $x_{Thymol} = 0.7$-$0.8$ (Fig. S1c,d), crystallization was avoided on cooling and occurred only during heating, at temperature well above the glass transition.

- poor glassformers: for thymol-rich systems, $x_{Thymol} \geq 0.9$ (Fig. S1a,b) crystallization occurred on cooling at 5 K.min$^{-1}$ but could be avoided by quenching at 200 K.min$^{-1}$.

- highly crystallizable systems: for menthol-rich systems, $x_{Thymol} \leq 0.2$ (Fig. S1i-k) crystallization occurred on cooling even during thermal quench at 200 K.min$^{-1}$. Note that the classification as 'highly crystallizable systems' is strictly valid for the experimental conditions used in the present study, but crystallization might be avoided, provided different experimental conditions are applied such as ultrafast cooling or nanoscale spatial confinement [30-32].

The temperatures of the different transitions measured by DSC (liquidus, crystallization on cooling, cold crystallization on heating and the glass transition) are summarized as filled symbols on the phase diagram illustrated in Fig. 1. The open symbols correspond to DS measurements and are discussed in the next section. The assignment of the different solid phases ($S_{Menthol}$, α, and β) as well as the two eutectic points (star symbols labelled $E_1$ and $E_2$) are indicated according to the comprehensive study by Alhadid *et al*. [23].

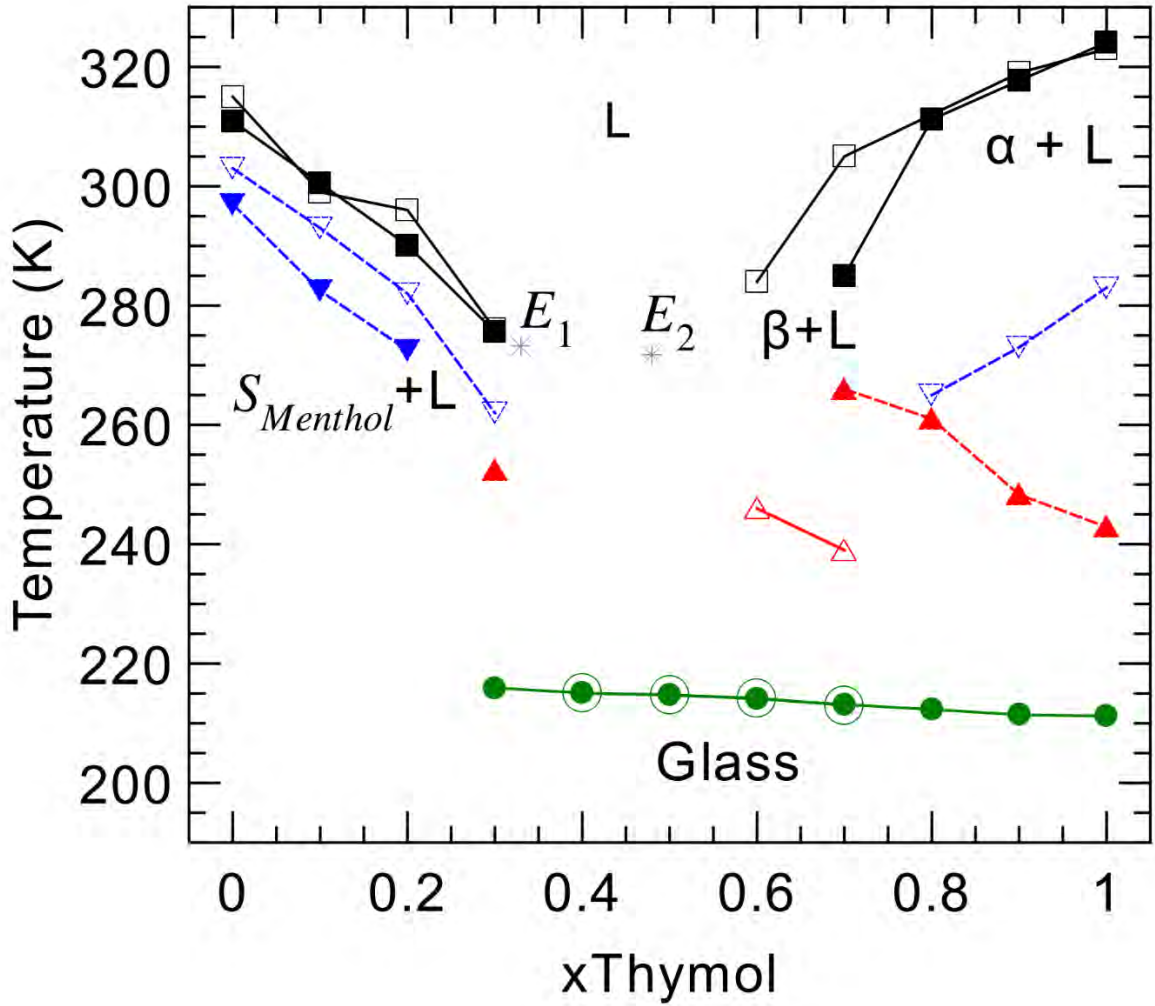

**FIG. 1**. Experimental solid–liquid and liquid-glass transitions for the binary thymol-menthol system measured by DSC (filled symbols) and dielectric spectroscopy (open symbols). Liquidus (black squares), crystallization temperature on cooling (blue downward triangles) and cold crystallization on heating (red upward triangles). The glass transition temperatures (green circles) correspond to entirely vitrified systems, i.e. with no partial crystallization (see text for experimental conditions). The two eutectic points denoted $E_1$ and $E_2$ (purple stars) and the assignment of the solid phases ($S_{Menthol}$, $\alpha$, and $\beta$) are based on Alhadid et al. [23].

The thermograms of the eight glassforming mixtures ($x_{\text{Thymol}} \geq 0.3$) in the region of the glass transition are illustrated in Fig. 2a. The glass transition is identified by a characteristic jump of the heat capacity that gradually shifts to lower temperature when the fraction of thymol increases. The glass transition temperature $T_g$ and the magnitude of the heat capacity jump at $\Delta C_p(T_g)$ are shown in Fig. 2b.

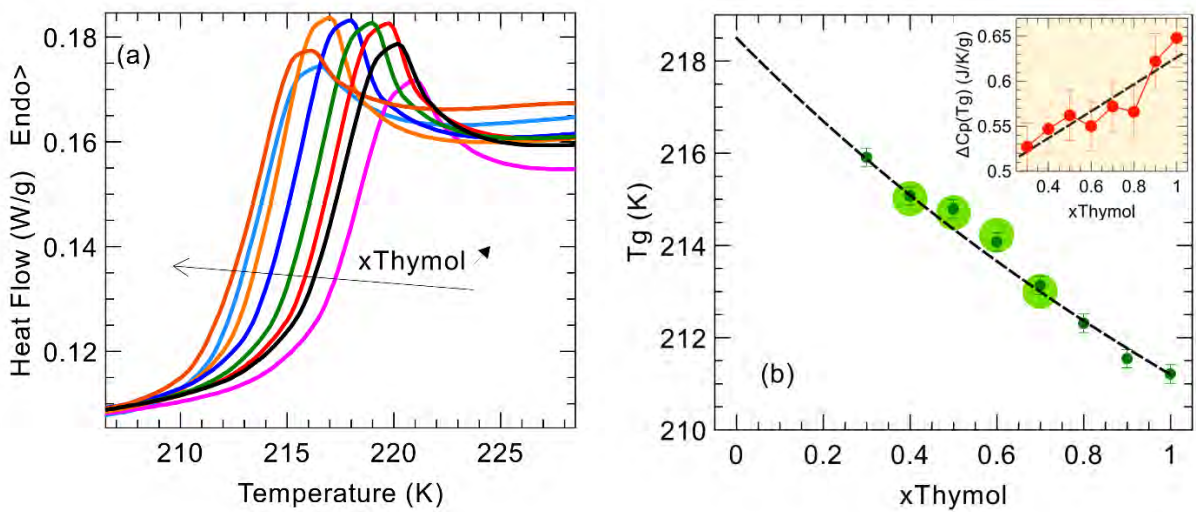

**FIG. 2**. (a) Thermograms measured during heating at 5°C.min$^{-1}$ in the region of the glass transition of entirely vitrified thymol-menthol mixtures for eight different values of the thymol molar fraction $x$ ranging from 0.3 to 1 from right to left (cf. arrow). (b) Dependence on the mixture composition of the glass transition temperature $T_g$ and (inset) the heat capacity jump at $T_g$. Values of $T_g$ evaluated from dielectric spectroscopy are added for four compositions (large symbols). Dashed line is fit with Gordon-Taylor model.

We compared the compositional dependence of the glass transition temperature with prediction based on the configurational entropy theory of the glass transition with ideal mixing approximation. According to the Gordon-Taylor theory, the glass transition of the mixture is given by eq. 1 with $K = \frac{\Delta C_{p2}(T_{g2})}{\Delta C_{p1}(T_{g1})}$, $x$ being the molar fraction of thymol, and 1, 2 indices standing for thymol and menthol, respectively.[33, 34]

$$T_g(x) = \frac{xT_{g1}+(1-x)KT_{g2}}{x+(1-x)K} \quad (1)$$

The parameter $K$ was not considered as an adjustable parameter, but determined from the experimental values of $\Delta C_p(T_g)$. To limit the effect of statistical fluctuations in the evaluation of $K$, we applied a linear regression, as indicated by the dashed line in the inset of Fig. 2b. The glass transition temperature of pure menthol being unknown, the Gordon-Taylor model was calculated using the values of the two extreme compositions ($x_{Thymol}$ = 0.3 and 1). The predicted temperatures are illustrated by the dashed curve in Fig. 2b. Interestingly, the values of the heat capacity jump ($\Delta C_p(T_g)$ = 0.48 J.K$^{-1}$.g$^{-1}$) and the glass transition temperature ($T_g$ = 218.5 K) of the highly crystallizable menthol could therefore be determined by extrapolation. The obtained value of $T_g$ is in good agreement with the previous prediction ($T_g$ = 218.8 K) obtained from a comparable extrapolation method [32]. A positive deviation of the compositional dependence of the glass transition temperature with respect to ideal mixing is observed at intermediate compositions ($x_{Thymol}$ = 0.5 and 0.6). The magnitude of this effect of non-ideality is however quite small, although significant within the experimental uncertainties. Additionally, this non-ideal variation of the calorimetric glass transition was also demonstrated by dielectric spectroscopy, as explained in the next section, by extrapolating the relaxation time to $10^2$s (cf. large filled circles in Fig. 2b). This positive deviation signifies that, on approaching $T_g$, the glassy dynamics of the DESs are slower than anticipated for ideal mixtures. This is consistent with the formation of preferential menthol-thymol H-bond interactions for near equimolar compositions that are expected to increase the solvent viscosity.[24]

*3.2 Dielectric spectroscopy – Phase behavior and dielectric strength*

While the calorimetric glass transition is usually associated to temperature at which the liquid dynamics is very slow (relaxation time of the order of $10^2$ s), the DS allows scrutinizing the liquid dynamics on a much wider range of timescales and thus at different temperatures. The complex dielectric function of the sample $\varepsilon^*(f) = \varepsilon'(f) - i\varepsilon''(f)$ was evaluated for the eleven DES samples where $f$ denotes the frequency of the electric field, $\varepsilon'$ and $\varepsilon''$ the real and loss part of the complex dielectric function and $i$ symbolizes the imaginary unit. They are illustrated as 3D plots as a function of the temperature $T$ and frequency $f$ for the equimolar DES mixture in Figs. 3a and 3b. Depending on the temperature and the frequency ranges, different contributions were observed in the instrumental frequency window. At high temperature and low frequency, the real part of the permittivity increases considerably (Fig. 3a). This is a well-known effect due to electrode polarization induced by moving ionic charges (possibly including impurities) that are eventually blocked and form a layer at the cell interface.

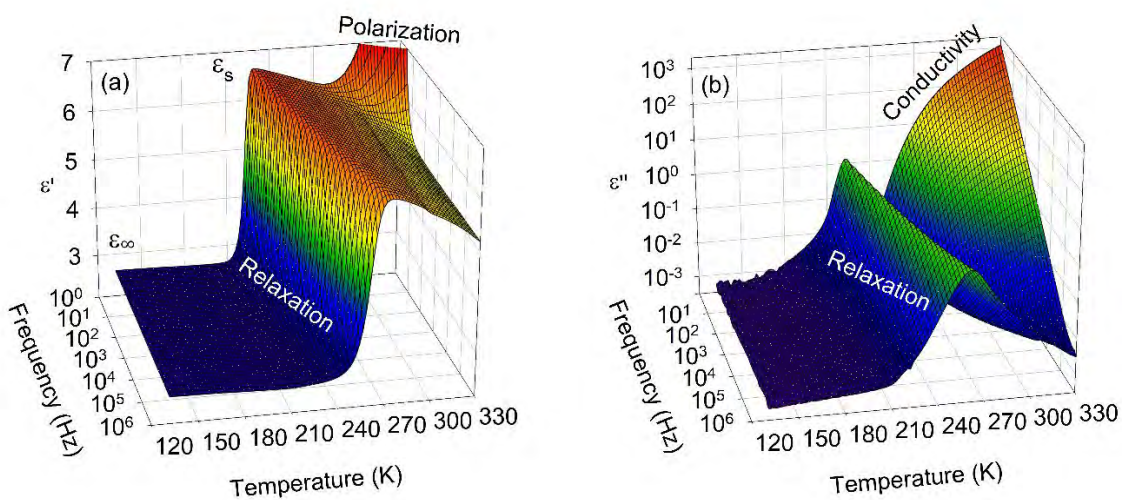

**FIG. 3**. 3D representation of the complex dielectric function measured for the equimolar thymol-menthol mixture as a function of the temperature and frequency. (a) Real part and (b) imaginary part.

Menthol-thymol being nonionic DESs, this contribution was rather limited and it could be easily separated from the physically interesting contributions. The most remarkable observed feature was the prominent dipolar relaxation process that is illustrated in Fig. 3a by the jump of ε'(*f*) from static permittivity $\varepsilon_s$ to high-frequency permittivity $\varepsilon_\infty$. Likewise, this relaxation process appears as a distinct peak in the loss part ε''(*f*). Conductivity additionally contributes to ε''(*f*) at high temperature and frequency.

For each sample, more than 150 spectra covering the frequency range (1Hz – 1Mhz) were acquired at varying temperatures by steps of 2K on cooling and heating. Isochronal representations of the real and imaginary parts of the complex dielectric function of all the studied menthol-thymol mixtures are illustrated in Figs. S2(a-v) for a limited selection of frequencies. Despite the relaxed temperature resolution of DS compared to DSC, this representation provides a direct insight on the occurrence of phase transitions and the nature of the thermodynamic state of the sample. Crystallization-melting of non-glassforming systems is obviously demonstrated by a hysteretic loop of the real part ε'(T), as illustrated for instance for pure thymol in Fig. S2a. Good glassformers are indicated by a sudden drop of ε'(T) on heating (cold crystallization) followed by a second jump at melting, as illustrated in Fig. S2g by the typical case $x_{Thymol}$ = 0.7. Finally, very good glassformers are demonstrated by a fully reversible temperature variation of ε'(T) and ε''(T), with a frequency dependent gradual variation of both quantities that indicates the crossing of the dipolar relaxation with the experimental dynamical window. This last behavior is nicely illustrated in Fig. S2k,l for the equimolar DES composition. Because no thermal quench was applied, and due to different sample geometry and much slower average scanning rate (ca. 0.3 K.min$^{-1}$), the samples studied by DS were more often subject to crystallization than when studied by DSC. The freezing/melting transition temperatures evaluated by DS are indicated in Fig. 1 by open symbols. There is a general agreement between both methods, although crystallization being a partly stochastic phenomenon, a quantitative agreement is not expected in this case. The DS study of the dynamics on the temperature range encompassing the supercooled liquid down to the glassy state could be achieved, without any crystallization, for four compositions $x_{Thymol}$ = 0.4-0.7.

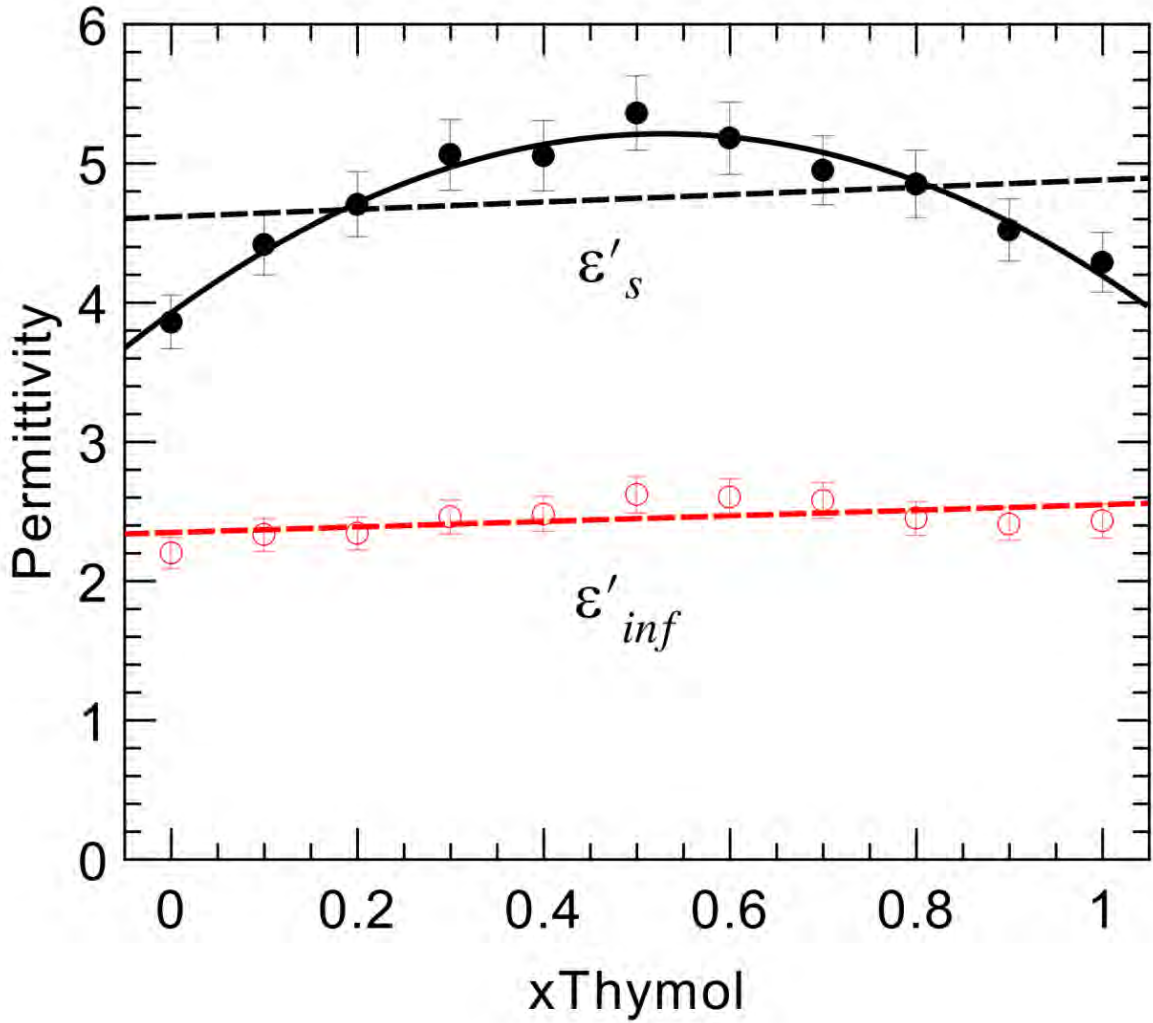

**FIG. 4**. The dependence on the composition of the thymol-menthol binary systems of the measured dielectric permittivity in the limits of low frequency $\varepsilon_s$ (black filled circles) and high frequency $\varepsilon_\infty$ (red open circles). The static permittivity corresponds to the temperature $T = 323$ K, in the liquid phase. Dashed lines and solid lines are fits with linear and quadratic functions, respectively.

Before addressing the dynamical properties of the glassforming mixtures, we first discuss the dielectric strength $\Delta\varepsilon = \varepsilon_s - \varepsilon_\infty$. It is a measure of the jump of $\varepsilon'(T)$ from the low temperature - low frequency limit to the high temperature - high frequency limit. As shown in Figs. S2, $\varepsilon_s$ varies with the temperature. It could be measured for all the eleven samples in the liquid state at a sufficiently high temperature $T$ = 323 K. At low temperature, in the crystalline phase or in the supercooled liquid, provided that $\omega\tau \gg 1$, all the $\varepsilon'(T)$ curves converge towards the barely temperature dependent limit $\varepsilon_\infty$. $\varepsilon_\infty$ rather obeys a flat linear compositional dependence, as illustrated by the dashed line in Fig. 4. The latter is expected, because $\varepsilon_\infty$ is associated to molecular electronic polarizability, and as such, it mostly depends on the molar concentration of each species. Minor deviations from a strict linear dependence are within the experimental uncertainties. Contrariwise, the static permittivity presents a significant dependence on the DES composition. It is properly described by an almost symmetric quadratic shape, with a maximum value 25% larger than for the pure systems, located close to the equimolar DES composition. The value of $\varepsilon_s$ emphasizes the correlations between the dipoles of different molecules present in the liquid. According to the Kirkwood−Fröhlich equation, when specific interactions such as H-bonds favor parallel alignment of the molecular dipoles, they lead to an enhanced effective dipole moment.[35] This phenomenon results in a larger dielectric strength. Contrariwise, antiparallel dipolar configurations reduce the value of the dielectric strength. In this framework, one may conclude that the H-bonded structures induced by the specific interactions between the unlike molecules promote parallel alignment of their dipoles. A similar phenomenon has been demonstrated for mixtures of monohydroxy alcohols.[36] Whereas ringlike low-dipole moment clusters prevail in the pure liquids, chainlike arrangements were adopted in the mixtures. This evolution of supramolecular structures with the composition, which was associated to non-ideal mixing effects, has been illustrated by the presence of a prepeak in the static structure factor. The relation between supramolecular H-bonded clusters and the prepeak in the static structure factor has been established for many other different H-bonded liquids, including alcohols and amines.[37-40] Since it results from the balance between H-bond interactions and geometric constraints due to steric effects, it is often observed for H-bonding molecules comprising a globular group such as phenyl (e.g. *m*-toluidine and *m*-cresol). The structural similarity of such molecules with menthol and thymol is obvious and it is worth noting that a prepeak in the static structure factor has been actually observed for the equimolar menthol-thymol DES mixture.[24]

*3.2 Dielectric spectroscopy – Dynamics of glassforming DES mixtures*

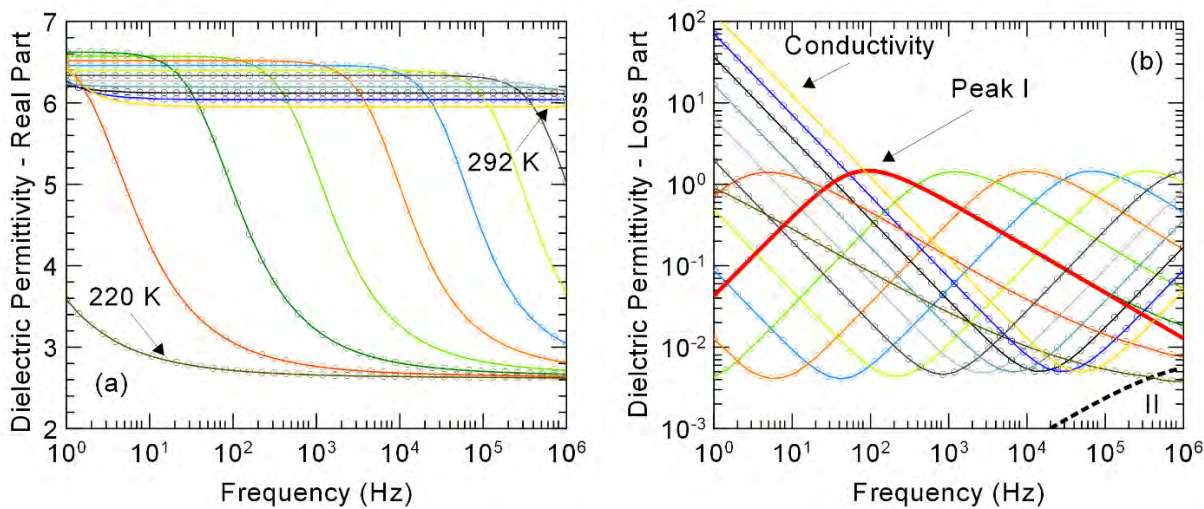

**FIG. 5**. The temperature dependence of (a) the real part and (b) the imaginary part of the complex dielectric function measured for the equimolar thymol-menthol mixture as a function of the frequency. For clarity, a limited number of temperatures were selected, varying by steps of 6 K in the range $T = 220 - 292$ K. Fitted functions (solid lines) are undistinguishable from the experimental data (circles). The two dipolar relaxations modeled by Havriliak–Negami functions,

peak I (red solid line) and peak II (black dashed line) are illustrated in panel (b) for the temperature 232 K.

We now discuss the dynamics of the DES for glassforming compositions $x_{\text{Thymol}} = 0.4$-$0.7$. The complex dielectric function was analyzed quantitatively at each temperature that was actually corresponding to a liquid phase by fitting a model including a Havriliak and Negami functions (HN-model),[41] and a dc-conductivity term according to eq. 2

$$\varepsilon^*(\omega) = \varepsilon_\infty + \frac{\Delta\varepsilon}{(1+(i\omega\tau_{HN})^{\alpha_{HN}})^{\beta_{HN}}} - i\frac{\sigma}{\omega\varepsilon_0} \qquad (2)$$

where $\omega = 2\pi f$, $\varepsilon_\infty$ is the sample permittivity in the limit of high frequency, $\Delta\varepsilon$ and $\tau_{HN}$ are respectively the dielectric strength and the HN-relaxation time of the mode. The spectra are illustrated in Fig. 5 for the equimolar composition and for a selection of temperatures. The spectra corresponding to all the measured samples are provided in Figs. S3a-v. $\sigma$ stands for the dc-conductivity of the sample and $\varepsilon_0$ the permittivity of vacuum. According to the formalism of the HN-model, the exponents $\alpha_{HN}$ and $\beta_{HN}$ ($0 < \alpha_{HN}$ ; $\alpha_{HN}\beta_{HN} \leq 1$) are fractional parameters describing, respectively, the symmetric and asymmetric broadening of the complex dielectric function with respect to the Debye one. [42] Their values varied only a little with temperature and composition, with $0.54 < \alpha_{HN} < 0.6$ and $\beta_{HN} > 0.94$, which is close to the Cole-Davidson equation.

For some temperatures, a small additional contribution which was generally overwhelmed by the main relaxation (peak I), appeared in the high frequency limit of the instrument. When needed, it was accounted for by an additional HN-model (peak II). The different contributions of the fitted model are illustrated in Fig. 5b at a single temperature. The overall model (solid lines in Figs. 5) provided an excellent fit to the data and was undistinguishable from the experimental points (circles). Due to the very small intensity of peak II (more than 2 magnitudes less intense than peak I) and its location at frequency above the spectral limit of the measurement, no reliable parameters could be obtained out of it.

The relaxation time associated to the maximum in the lost part of the complex dielectric function frequency was computed according to [43]

$$\tau = \tau_{HN} \sin\left(\frac{\pi \alpha_{HN}}{2+2\beta_{HN}}\right)^{-1/\alpha_{HN}} \sin\left(\frac{\pi \alpha_{HN} \beta_{HN}}{2+2\beta_{HN}}\right)^{1/\alpha_{HN}} \qquad (3)$$

The Arrhenius plots of the relaxation time presented in Fig. 6 exhibit large deviations from an Arrhenius behavior. The increase of the apparent activation energy on approaching the glass transition is a salient feature of supercooled liquids, and it has been also observed for hydrophilic DESs based on choline chloride and their aqueous mixtures.[14, 28, 29] The classical Vogel–Fulcher–Tammann (VFT) law (Eq. 4) was fitted to the dipolar relaxation time,

$$\tau = \tau_0 exp\left(\frac{DT_0}{T-T_0}\right) \qquad (4)$$

where the pre-exponential factor $\tau_0$ is a characteristic microscopic time, $D$ and $T_0$ are the strength parameter and the Vogel-Fulcher-Tammann temperature. The VFT fits (solid line in Fig. 6) were used to extrapolate the relaxation time, and they provided a complementary estimation of the glass transition defined by $T_g(\tau = 10^2$ s). A very good agreement was obtained with the calorimetric glass transition temperatures that are indicated by downward triangles in Fig. 6. Direct comparisons between the values of $T_g$ evaluated by DSC and by DS are also provided in Fig.1 and Fig. 2b.

From Fig. 6, a systematic acceleration of the dipolar relaxation on increasing the mole fraction of thymol is observed. This is consistent with the decrease of $T_g$ shown in Fig. 2b. This acceleration effect also depends on the temperature range and it is enhanced at low temperature. For instance, the relaxation times of the two extreme compositions ($x_{Thymol}$ = 0.4, and 0.7) differ by a factor 2 at room temperature, but they differ by more than a factor 5 on approaching the glass transition. This observation can be paralleled with the predominance of H-bond interactions and non-ideal effects on cooling evidenced by Schaeffer *et al*. [24]. The temperature dependence of the dynamics of the different DES mixtures can be quantified by the strength parameter $D$. The smaller the value of $D$, the greater the deviation from Arrhenius behavior. It is also commonly rewritten in terms of the fragility index $m$, that categorizes glass-forming liquids as fragile (with marked super-Arrhenian character) or strong systems (obeying the Arrhenius law).[44] We obtained values of the fragility index of menthol-thymol DES in the range $m$ = 77 to $m$ = 86. They are characteristic of fragile liquids. As a comparison, values in the range from $m$ = 60 to $m$ = 40 were reported for the neat ionic DES ethaline and for different hydration levels thereof, which were thus classified as intermediate liquids.[14, 29] As illustrated in Fig. 6a, the fragility index $m$ exhibits a systematic increase when increasing the fraction of thymol. A different conclusion was derived for water-ethaline systems: indeed, while adding water also led to an acceleration of the dynamics, like for thymol in menthol-thymol DES, it had the opposite effect on the fragility.[14] This highlights the complexity of the physico-chemical properties of DESs, which result from a delicate balance of different intermolecular interactions.

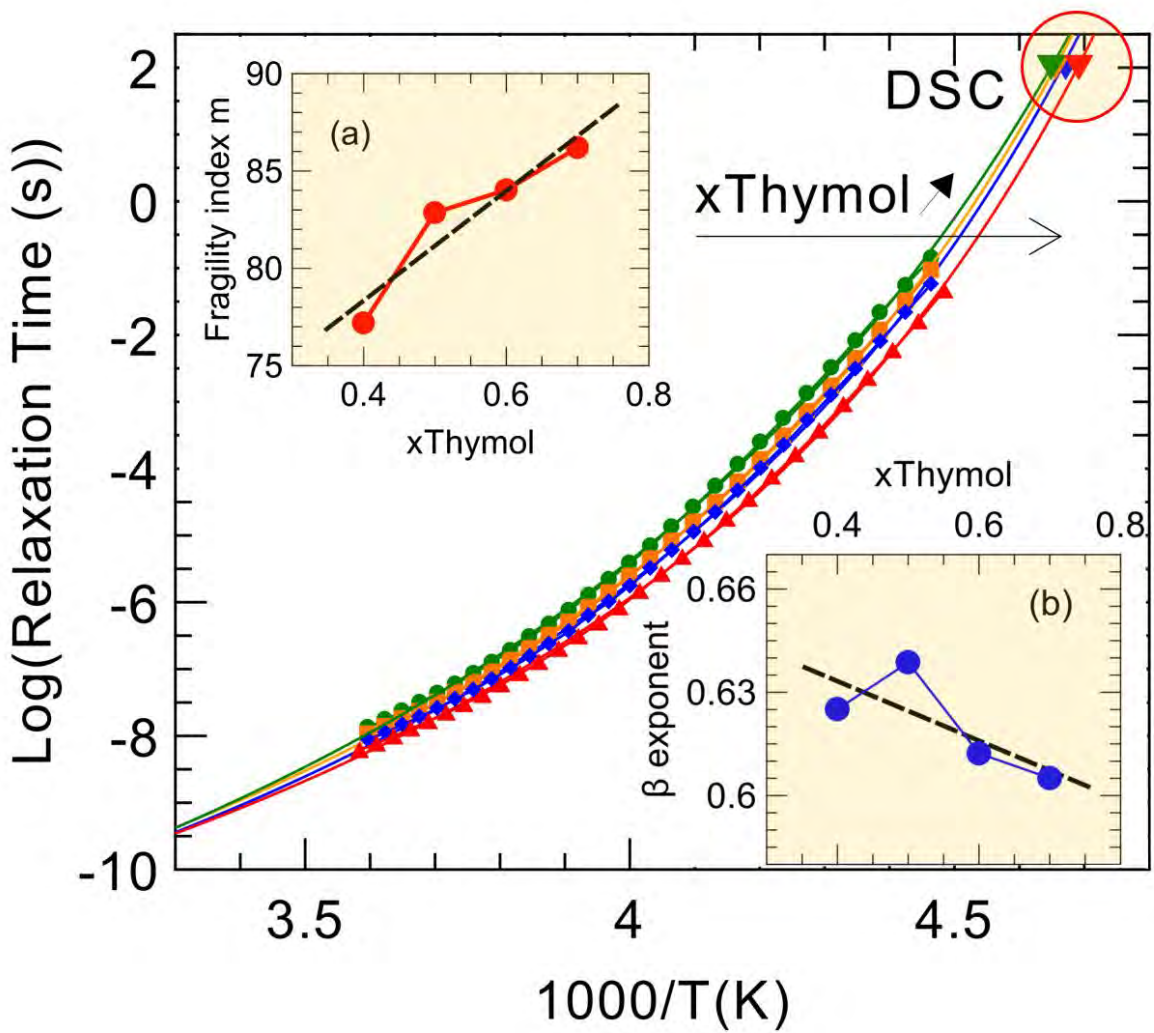

**FIG. 6**. Arrhenius representation of the temperature dependence of the relaxation time of glassforming thymol-menthol binary liquids for thymol molar fraction $x_{Thymol}$ = 0.4 (green circles), 0.5 (orange squares), 0.6 (blue diamonds), and 0.7 (red triangles). Solid lines are fits with a VTF law. The glass transition temperatures measured by DSC are added at the corresponding relaxation time $\tau(Tg) = 10^2$ s (downward triangles). Inset (a): the fragility index $m$ determined from the VTF

fits of the relaxation times. Inset (b): the Kohlrausch stretching exponent $\beta$ determined from the fits of the dielectric relaxation functions. Dashed lines are linear regressions.

Another point of interest concerns the deviation of the dipolar relaxation function from a pure Debye process. It implies that, in the time domain, the relaxation function exhibits a non-exponential character that is commonly represented by a stretched exponential function denoted as the Kohlrausch-William-Watts equation. The stretching exponents $\beta$ were deduced from the HN parameters obtained in the frequency domain, applying the numerical ansatz $= (\alpha_{HN}\beta_{HN})^{\frac{1}{1.23}}$. [45] They are significantly smaller than one, as illustrated in Fig. 6b, taking values in the range 0.6 - 0.64, which are typical for the main structural relaxation of fragile glassforming liquids. This confirms that the observed dipolar process actually corresponds to the main liquid relaxation, as already supported by the agreement between $T_g$'s deduced by DSC and DS, and it rules out its hypothetical attribution to the Debye-like mode often observed for monohydroxy alcohols.[46-48] A tendency of modest decrease of $\beta$ with increasing $x_{Thymol}$ is also perceived in Fig. 6b. This compositional dependence of $\beta$ can be correlated to the increase in the fragility index $m$, thus following the relationship between fragility and non-exponentiality that has been established for numerous glass-forming systems.[44] This indicates that, in this range of compositions, the dipolar relaxation of the DES is increasingly associated to the presence of dynamic heterogeneities.[49] It should be emphasized however that dynamic heterogeneities are salient features of deeply supercooled systems, and as such they are also often observed for pure liquids. [49] Therefore, they cannot be taken as direct evidence for the hypothesized tendency of DES to actually phase separate on a mesoscopic or larger scale.

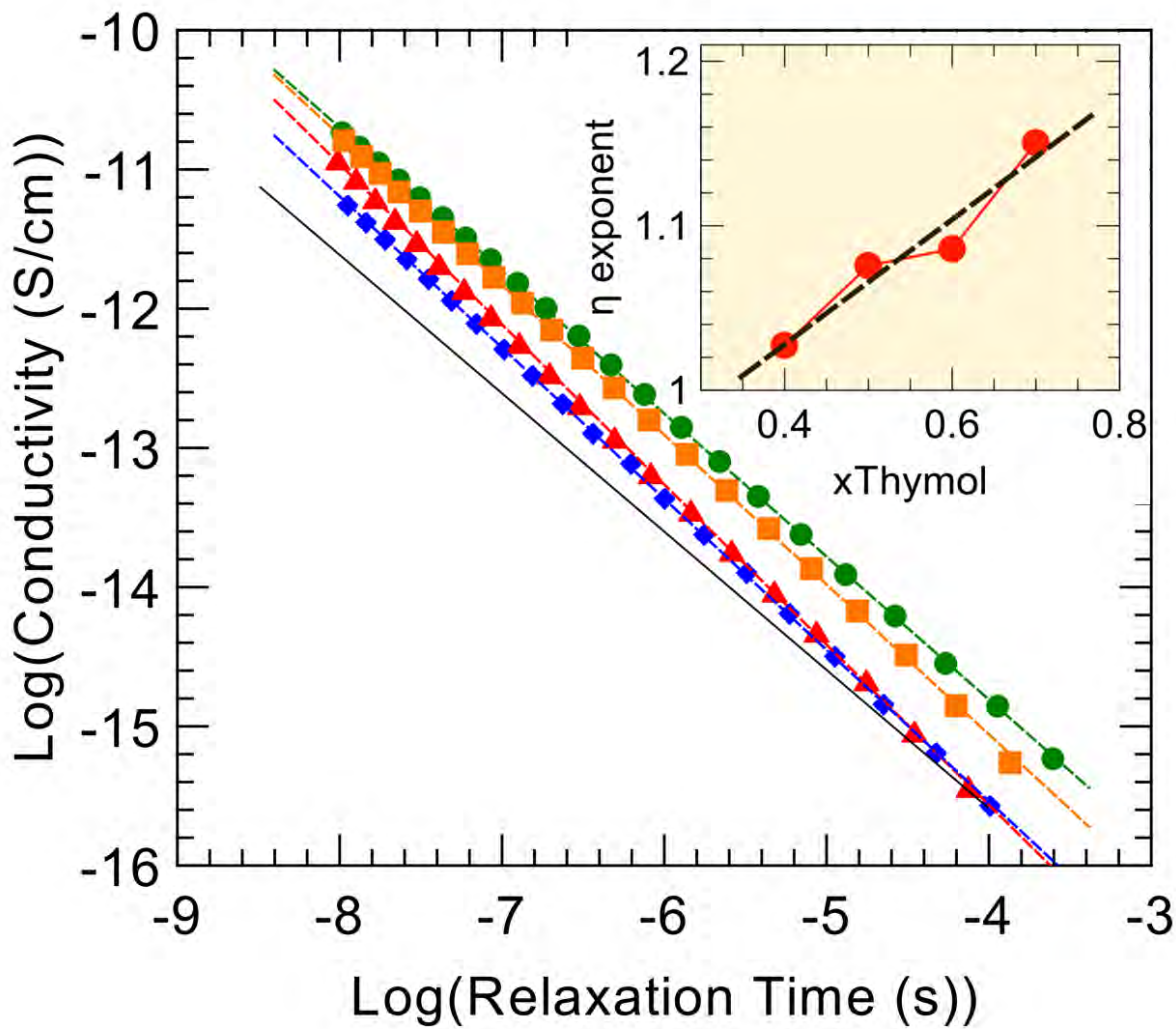

**FIG. 7**. Dependence of the dc-conductivities σ on the reorientational relaxation times τ of glassforming thymol-menthol binary liquids for thymol molar fraction $x$ = 0.4 (green circles), 0.5 (orange squares), 0.6 (blue diamonds), and 0.7 (red triangles). The dashed lines are fits with power law $\sigma \propto \tau^{-\eta}$ while the solid line with slope -1 indicates inverse proportionality $\sigma \propto \tau^{-1}$. Inset: the dependence of the exponent η on the composition of the thymol-menthol binary liquids. Dashed line is a linear regression.

A classical indirect indication of spatial dynamic heterogeneities is the decoupling between translational and rotational molecular diffusion.[49] This phenomenon stems from the fact that, in a heterogeneous system, the experimental evaluation of the rotation and the translation are averaged differently over the heterogeneity. This leads to an apparent violation of the Stokes-Einstein and Stokes-Einstein-Debye relationships that relate both dynamics to viscosity at the hydrodynamic limit. This hypothesis is questioned in Fig. 7 where the reorientational relaxation times $\tau$ is represented as a function of the dc-conductivities $\sigma$. A deviation from inverse proportionality $\sigma \propto \tau^{-1}$ is observed. It was quantified by fitting a power law $\sigma \propto \tau^{-\eta}$ (dashed line in Fig. 7). The obtained values of the exponent $\eta$ illustrated in inset of Fig. 7 gradually increases on increasing $x_{Thymol}$. Again, this points to the development of spatial dynamic heterogeneities in the glassforming menthol-thymol DES in this range of compositions. For comparison, in the case of the prototypical ionic DES ethaline, no translation-rotation decoupling was observed, even for moderately hydrated systems, until such large level of water content led to a truly biphasic system.[14]

## Conclusions

Type V DESs have been very recently introduced as a novel class of DES. Due to their nonionic character, their hydrophobic nature and their biocompatibility, they currently constitute one of the most exciting sources of inspiration for the development of new applications of DESs. However, the field is still in its infancy and fundamental knowledge of this new family of solvents remains incomplete. Even for the prototypical case of menthol-thymol, the comprehensive studies of its thermodynamic behavior and H-bonds structure lack equivalent investigations of its dynamics.

Combining DSC and DS, the present study provides an extended description of the dynamics of menthol-thymol mixtures for eleven different compositions including the two pure constituents, and for a timescale spanning from $10^{-6}$ s to $10^2$ s. Several measured properties indicate a possible link between non-ideality, as induced by the preferential H-bond interactions between unlike molecules, and important features of the molecular dynamics.

The quadratic compositional dependence of the dielectric strength with a maximum centered at about the equimolar DES definitely supports that the equilibrium between different populations of supramolecular H-bonded clusters is displaced towards more linear and less cyclic structures by thymol-menthol interactions. These preferential interactions are also highlighted by a resulting positive deviation of the glass transition, as determined by DSC and DS, with respect to predictions based on the ideal mixing approximation.

The dipolar relaxation dynamics could be studied for a limited range of compositions, where the DESs exhibit excellent glassforming tendency $x_{Thymol}$ = 0.4-0.7. They exhibit all the salient features of glassforming systems, and they can be classified as fragile liquids. The non-Arrhenius temperature dependence of the relaxation time quantified by the index of fragility $m$, the non-Debye character of the dipolar relaxation function quantified by the Kohlraush exponent $\beta$, and the translation-rotation decoupling all indicate the existence of spatial dynamic heterogeneities. In this limited range of compositions, their impact on the dynamics presents a systematic, though modest, increase with increasing of $x_{Thymol}$.

These conclusions emerge from an unprecedented extensive investigation of the dynamics of a type V DES, and their relationship with H-bonds and non-ideality. However, from the present study of the single prototypical menthol-thymol system, we cannot make a definite distinction between universal and specific features of type V DESs. Surely, the search for general conclusions about the behavior of this promising new class of DES will inspire further work in the near future.

## Author statement

Claire D'Hondt: Investigation.

Denis Morineau: Conceptualization, Investigation, Writing – Original, Draft, Supervision, Funding acquisition.

## Declaration of competing interest

The authors declare that they have no known competing financial interests or personal relationships that could have appeared to influence the work reported in this paper.

## Acknowledgments

Support from Rennes Metropole and Europe (FEDER Fund – CPER PRINT$_2$TAN). The authors are grateful to the CNRS – network SolVATE (GDR 2035) for financial support and fruitful discussions.

## References

[1] A. Abbott, G. Capper, D. Davies, R. Rasheed, V. Tambyrajah, Novel solvent properties of choline chloride/urea mixtures, Chemical Communications, (2003) 70-71.

[2] A. Abbott, D. Boothby, G. Capper, D. Davies, R. Rasheed, Deep eutectic solvents formed between choline chloride and carboxylic acids: Versatile alternatives to ionic liquids, Journal of the American Chemical Society, 126 (2004) 9142-9147.

[3] Q. Zhang, K. Vigier, S. Royer, F. Jerome, Deep eutectic solvents: syntheses, properties and applications, Chemical Society Reviews, 41 (2012) 7108-7146.

[4] E. Smith, A. Abbott, K. Ryder, Deep Eutectic Solvents (DESs) and Their Applications, Chemical Reviews, 114 (2014) 11060-11082.

[5] X. Ge, C. Gu, X. Wang, J. Tu, Deep eutectic solvents (DESs)-derived advanced functional materials for energy and environmental applications: challenges, opportunities, and future vision, Journal of Materials Chemistry a, 5 (2017) 8209-8229.

[6] B.B. Hansen, S. Spittle, B. Chen, D. Poe, Y. Zhang, J.M. Klein, A. Horton, L. Adhikari, T. Zelovich, B.W. Doherty, B. Gurkan, E.J. Maginn, A. Ragauskas, M. Dadmun, T.A. Zawodzinski, G.A. Baker, M.E. Tuckerman, R.F. Savinell, J.R. Sangoro, Deep Eutectic Solvents: A Review of Fundamentals and Applications, Chemical Reviews, 121 (2021) 1232-1285.

[7] L. Kollau, M. Vis, A. van den Bruinhorst, A. Esteves, R. Tuinier, Quantification of the liquid window of deep eutectic solvents, Chemical Communications, 54 (2018) 13351-13354.

[8] M. Martins, S. Pinho, J. Coutinho, Insights into the Nature of Eutectic and Deep Eutectic Mixtures, Journal of Solution Chemistry, 48 (2019) 962-982.

[9] S. Kaur, A. Malik, H.K. Kashyap, Anatomy of Microscopic Structure of Ethaline Deep Eutectic Solvent Decoded through Molecular Dynamics Simulations, Journal of Physical Chemistry B, 123 (2019) 8291-8299.

[10] S. Kaur, A. Gupta, H.K. Kashyap, How Hydration Affects the Microscopic Structural Morphology in a Deep Eutectic Solvent, Journal of Physical Chemistry B, 124 (2020) 2230-2237.

[11] L. Percevault, A. Jani, T. Sohier, L. Noirez, L. Paquin, F. Gauffre, D. Morineau, Do Deep Eutectic Solvents Form Uniform Mixtures Beyond Molecular Microheterogeneities?, Journal of Physical Chemistry B, 124 (2020) 9126-9135.

[12] D.O. Abranches, M.A.R. Martins, L.P. Silva, N. Schaeffer, S.P. Pinho, J.A.P. Coutinho, Phenolic hydrogen bond donors in the formation of non-ionic deep eutectic solvents: the quest for type V DES, Chemical Communications, 55 (2019) 10253-10256.

[13] A. Jani, T. Sohier, D. Morineau, Phase behavior of aqueous solutions of ethaline deep eutectic solvent, Journal of Molecular Liquids, 304 (2020) 112701-112706.

[14] A. Jani, B. Malfait, D. Morineau, On the coupling between ionic conduction and dipolar relaxation in deep eutectic solvents: Influence of hydration and glassy dynamics, Journal of Chemical Physics, 154 (2021) 164508.

[15] B. Malfait, A. Jani, D. Morineau, Confining deep eutectic solvents in nanopores: Insight into thermodynamics and chemical activity, Journal of Molecular Liquids, 349 (2022) 118488.

[16] D. van Osch, C. Dietz, J. van Spronsen, M.C. Kroon, F. Gallucci, M.V. Annaland, R. Tuinier, A Search for Natural Hydrophobic Deep Eutectic Solvents Based on Natural Components, Acs Sustainable Chemistry & Engineering, 7 (2019) 2933-2942.

[17] D. Rodriguez-Llorente, A. Canada-Barcala, S. Alvarez-Torrellas, V.I. Agueda, J. Garcia, M. Larriba, A Review of the Use of Eutectic Solvents, Terpenes and Terpenoids in Liquid-liquid Extraction Processes, Processes, 8 (2020) 1220.

[18] O.G. Sas, L. Villar, A. Dominguez, B. Gonzalez, E.A. Macedo, Hydrophobic deep eutectic solvents as extraction agents of nitrophenolic pollutants from aqueous systems, Environmental Technology & Innovation, 25 (2022) 102170.

[19] M.A.R. Martins, L.P. Silva, N. Schaeffer, D.O. Abranches, G.J. Maximo, S.P. Pinho, J.A.P. Coutinho, Greener Terpene-Terpene Eutectic Mixtures as Hydrophobic Solvents, Acs Sustainable Chemistry & Engineering, 7 (2019) 17414-17423.

[20] A. Alhadid, L. Mokrushina, M. Minceva, Formation of glassy phases and polymorphism in deep eutectic solvents, Journal of Molecular Liquids, 314 (2020) 113667.

[21] D.O. Abranches, R.O. Martins, L.P. Silva, M.A.R. Martins, S.P. Pinho, J.A.P. Coutinho, Liquefying Compounds by Forming Deep Eutectic Solvents: A Case Study for Organic Acids and Alcohols, Journal of Physical Chemistry B, 124 (2020) 4174-4184.

[22] A. Alhadid, L. Mokrushina, M. Minceva, Influence of the Molecular Structure of Constituents and Liquid Phase Non-Ideality on the Viscosity of Deep Eutectic Solvents, Molecules, 26 (2021) 4208.

[23] A. Alhadid, C. Jandl, L. Mokrushina, M. Minceva, Experimental Investigation and Modeling of Cocrystal Formation in L-Menthol/Thymol Eutectic System, Crystal Growth & Design, 21 (2021) 6083-6091.

[24] N. Schaeffer, D.O. Abranches, L.P. Silva, M.A.R. Martins, P.J. Carvalho, O. Russina, A. Triolo, L. Paccou, Y. Guinet, A. Hedoux, J.A.P. Coutinho, Non-Ideality in Thymol plus Menthol Type V Deep Eutectic Solvents, Acs Sustainable Chemistry & Engineering, 9 (2021) 2203-2211.

[25] D.O. Abranches, J.A.P. Coutinho, Type V deep eutectic solvents: Design and applications, Current Opinion in Green and Sustainable Chemistry, 35 (2022) 100612.

[26] D.K. Panda, B.L. Bhargava, Molecular dynamics investigation of non-ionic deep eutectic solvents, Journal of Molecular Graphics & Modelling, 113 (2022) 108152.

[27] L. Zamora, C. Benito, A. Gutierrez, R. Alcalde, N. Alomari, A. Al Bodour, M. Atilhan, S. Aparicio, Nanostructuring and macroscopic behavior of type V deep eutectic solvents based on monoterpenoids, Physical Chemistry Chemical Physics, 24 (2021) 512-531.

[28] D. Reuter, P. Munzner, C. Gainaru, P. Lunkenheimer, A. Loidl, R. Böhmer, Translational and reorientational dynamics in deep eutectic solvents, Journal of Chemical Physics, 154 (2021) 154501.

[29] D. Reuter, C. Binder, P. Lunkenheimer, A. Loidl, Ionic conductivity of deep eutectic solvents: the role of orientational dynamics and glassy freezing, Physical Chemistry Chemical Physics, 21 (2019) 6801-6809.

[30] E. Zhuravlev, J. Jiang, D.S. Zhou, R. Androsch, C. Schick, Extending Cooling Rate Performance of Fast Scanning Chip Calorimetry by Liquid Droplet Cooling, Applied Sciences-Basel, 11 (2021) 3813.

[31] Y.D. Xia, G. Dosseh, D. Morineau, C. Alba-Simionesco, Phase Diagram and Glass Transition of Confined Benzene, Journal of Physical Chemistry B, 110 (2006) 19735-19744.

[32] T. Cordeiro, C. Castineira, D. Mendes, F. Danede, J. Sotomayor, I.M. Fonseca, M.G. da Silva, A. Paiva, S. Barreiros, M.M. Cardoso, M.T. Viciosa, N.T. Correia, M. Dionisio, Stabilizing Unstable Amorphous Menthol through Inclusion in Mesoporous Silica Hosts, Molecular Pharmaceutics, 14 (2017) 3164-3177.

[33] J.M. Gordon, G.B. Rouse, J.H. Gibbs, W.M. Risen, The composition dependence of glass transition properties, J. Chem. Phys., 66 (1977) 4971.

[34] M. Gordon, J.S. Taylor, Ideal copolymers and the 2nd-order transitions of synthetic

rubbers .1. Non-crystalline copolymers, Journal of Applied Chemistry, 2 (1952) 493-500.

[35] H. Fröhlich, General theory of the static dielectric constant, Transactions of the Faraday Society, 44 (1948) 238-243.

[36] S.P. Bierwirth, T. Buning, C. Gainaru, C. Sternemann, M. Tolan, R. Böhmer, Supramolecular x-ray signature of susceptibility amplification in hydrogen-bonded liquids, Physical Review E, 90 (2014) 052807.

[37] D. Morineau, C. Alba-Simionesco, Hydrogen-bond-induced clustering in the fragile glass-forming liquid m-toluidine: Experiments and simulations, Journal of Chemical Physics, 109 (1998) 8494-8503.

[38] D. Morineau, C. Alba-Simionesco, M.C. Bellissent-Funel, M.F. Lauthie, Experimental indication of structural heterogeneities in fragile hydrogen-bonded liquids, Europhysics Letters, 43 (1998) 195-200.

[39] D. Morineau, C. Alba-Simionesco, Does Molecular Self-Association Survive in Nanochannels?, Journal of Physical Chemistry Letters, 1 (2010) 1155-1159.

[40] A.R.A. Hamid, R. Lefort, Y. Lechaux, A. Moreac, A. Ghoufi, C. Alba-Simionesco, D. Morineau, Solvation Effects on Self-Association and Segregation Processes in tert-Butanol-Aprotic Solvent Binary Mixtures, Journal of Physical Chemistry B, 117 (2013) 10221-10230.

[41] S. Havriliak, S. Negami, A complex plane analysis of alpha-dispersions in some polymer systems, Journal of Polymer Science Part C-Polymer Symposium, (1966) 99-117.

[42] F. Kremer, A. Schönhals, Broadband Dielectric Spectroscopy, Springer, Berlin (2002).

[43] R. Diaz-Calleja, Comment on the maximum in the loss permittivity for the Havriliak-Negami equation, Macromolecules, 33 (2000) 8924-8924.

[44] R. Böhmer, K.L. Ngai, C.A. Angell, D.J. Plazek, Nonexponential relaxations in strong and fragile glass formers, Journal of Chemical Physics, 99 (1993) 4201-4209.

[45] F. Alvarez, A. Alegria, J. Colmenero, Relationship between the time-domain Kohlrausch-Williams-Watts and frequency-domain Havriliak-Negami relaxation functions, Physical Review B, 44 (1991) 7306-7312.

[46] L.M. Wang, S. Shahriari, R. Richert, Diluent effects on the Debye-type dielectric relaxation in viscous monohydroxy alcohols, Journal of Physical Chemistry B, 109 (2005) 23255-23262.

[47] M. Preuss, C. Gainaru, T. Hecksher, S. Bauer, J.C. Dyre, R. Richert, R. Böhmer, Experimental studies of Debye-like process and structural relaxation in mixtures of 2-ethyl-1-hexanol and 2-ethyl-1-hexyl bromide, Journal of Chemical Physics, 137 (2012) 144502.

[48] T. El Goresy, R. Böhmer, Diluting the hydrogen bonds in viscous solutions of n-butanol with n-bromobutane: A dielectric study, Journal of Chemical Physics, 128 (2008) 154520.

[49] M.D. Ediger, Spatially heterogeneous dynamics in supercooled liquids, Annual Review of Physical Chemistry, 51 (2000) 99-128.

# Supporting Information

# Dynamics of type V menthol-thymol deep eutectic solvents: Do they reveal non-ideality?

Claire D'Hondt, Denis Morineau [*]

[†]Institute of Physics of Rennes, CNRS-University of Rennes 1, UMR 6251, F-35042 Rennes, France

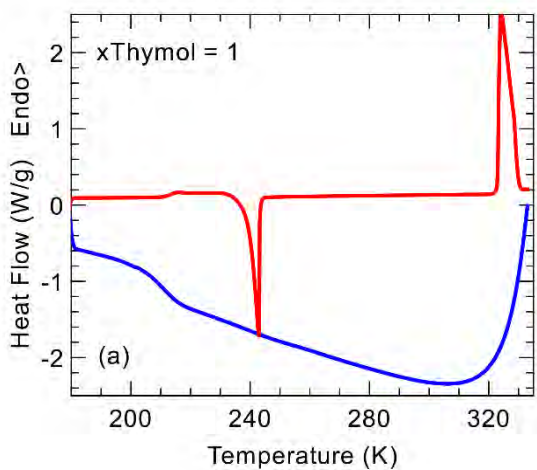
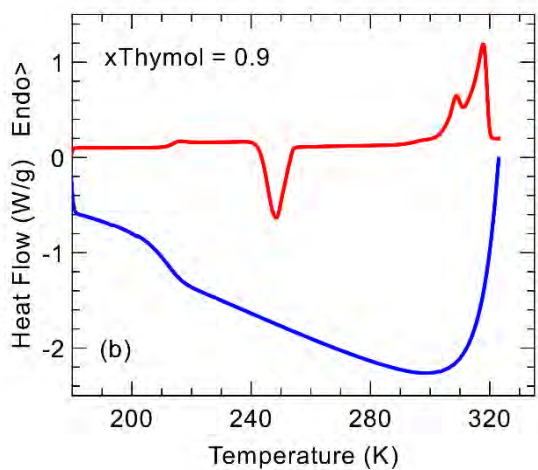
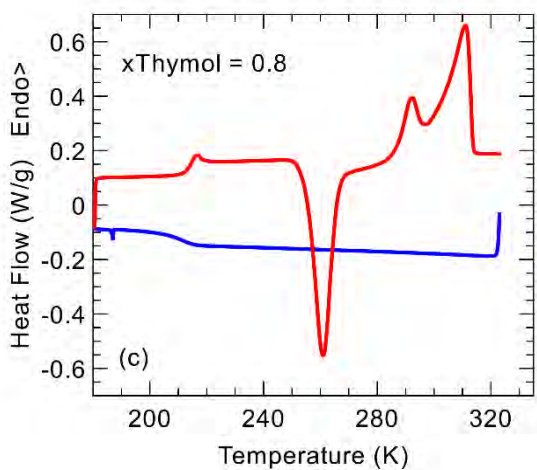
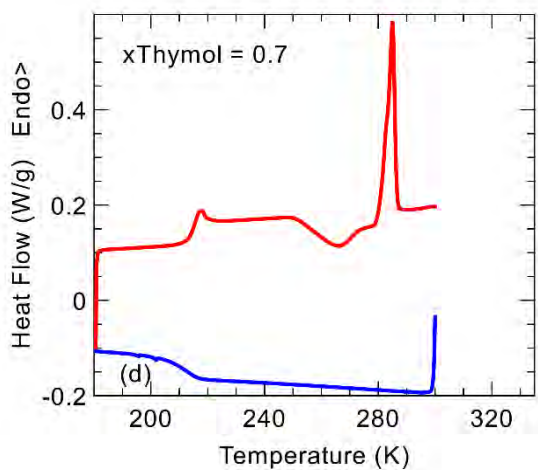
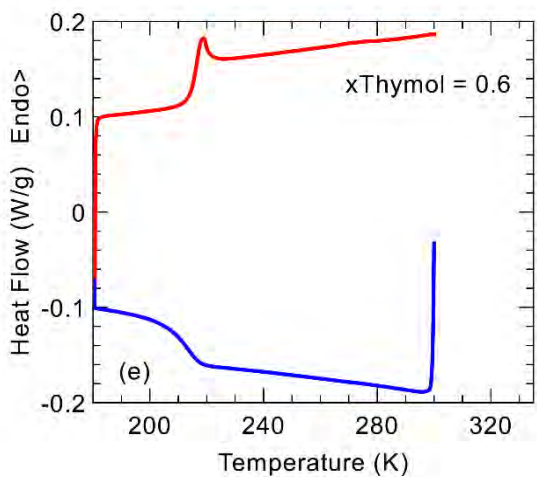
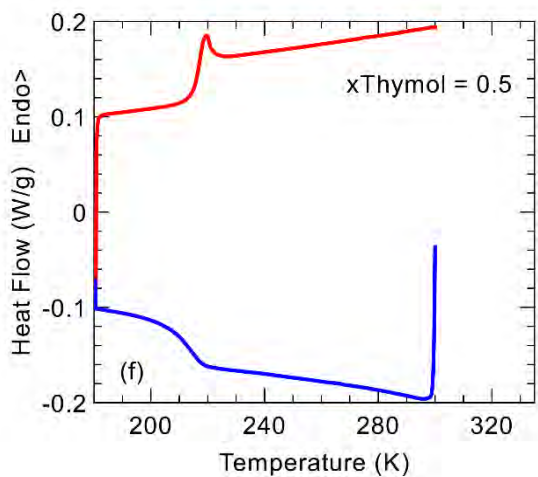

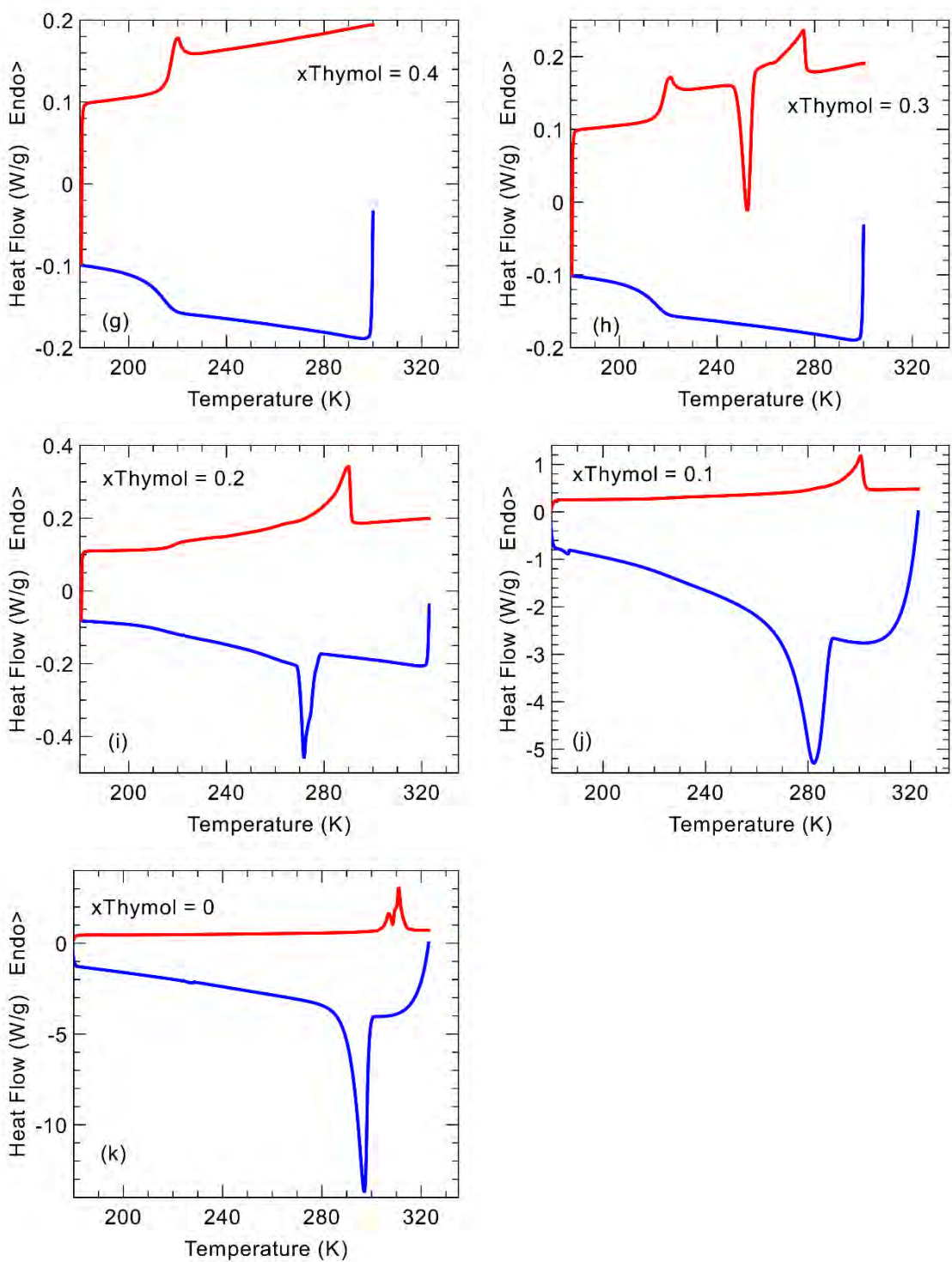

**Fig. S1.** Thermograms of menthol-thymol mixtures for different values of the molar fraction of thymol going from pure thymol (upper panel) to pure menthol (lower panel). Lower curves (blue) correspond to cooling branches, and upper curves (red) correspond to heating branches. Scanning rates are 5 K.min$^{-1}$, except for $x_{Thymol}$ = 1, 0.9, 0.1, and 0 (panels a, b, j, and k), where fast cooling rates of 200 K.min$^{-1}$ are applied.

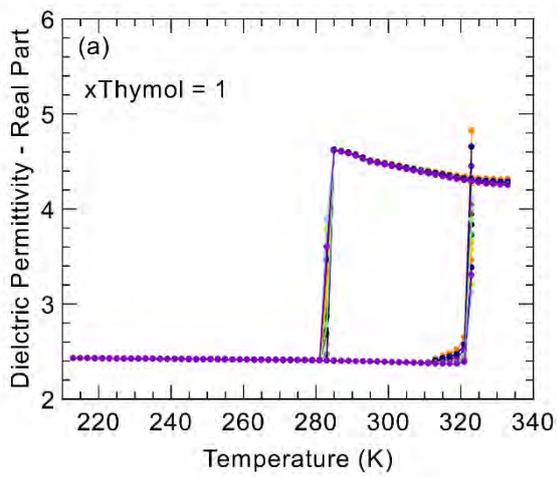
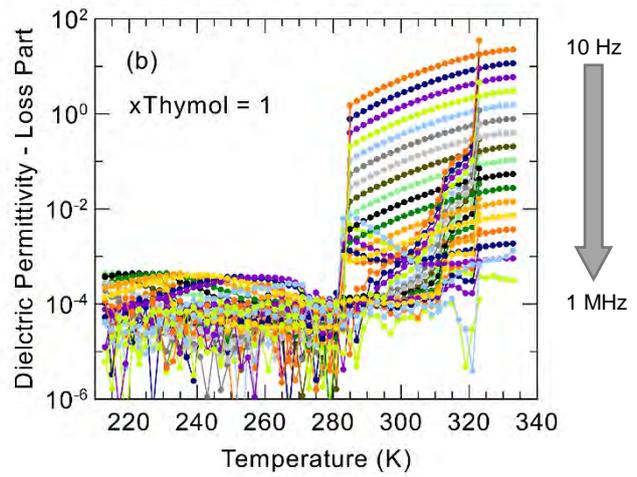
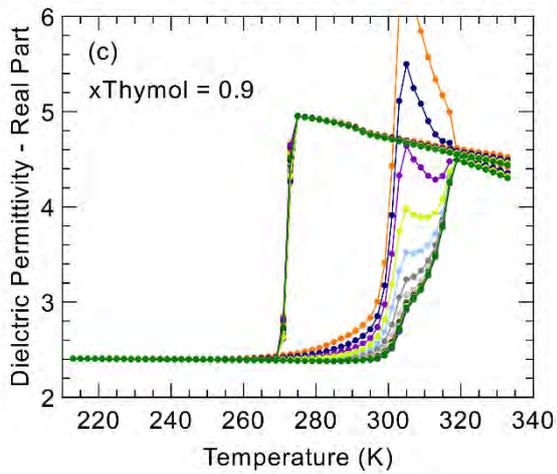
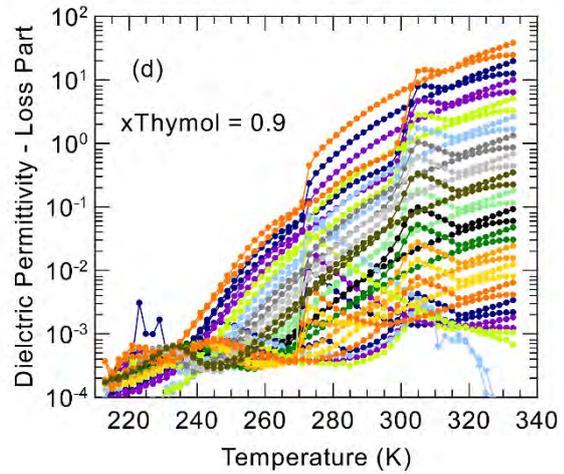
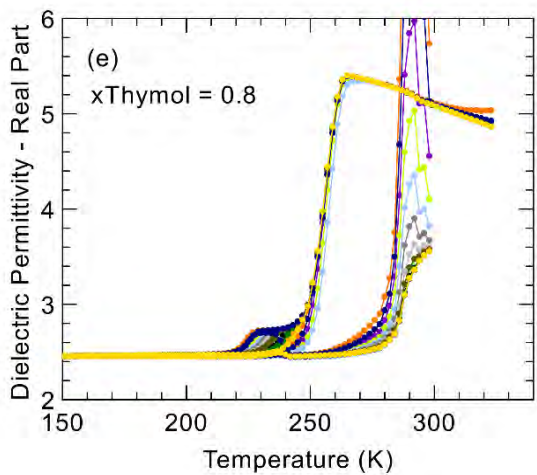
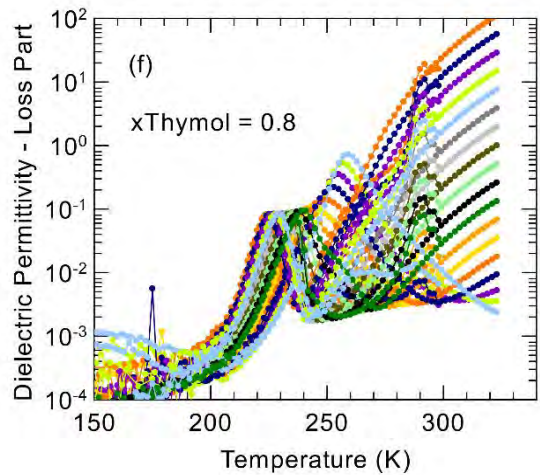

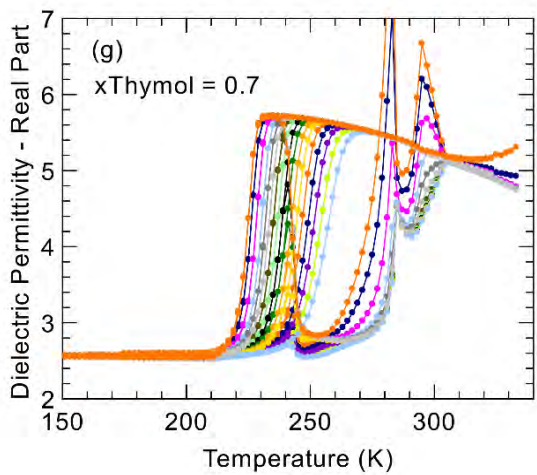
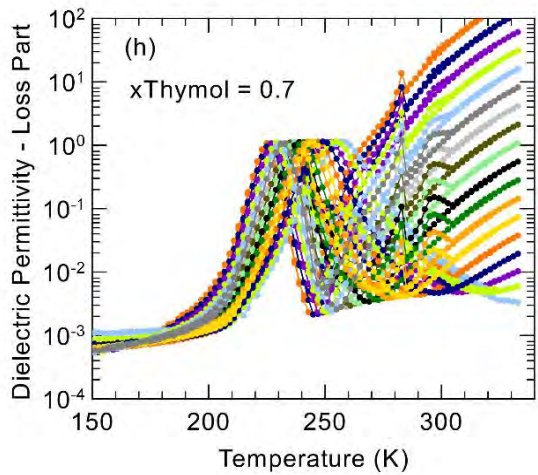
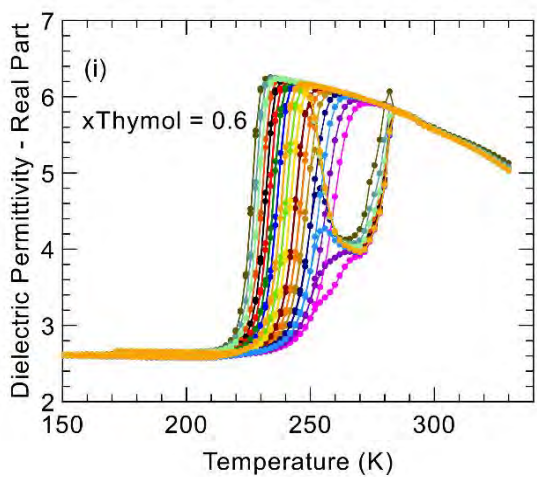
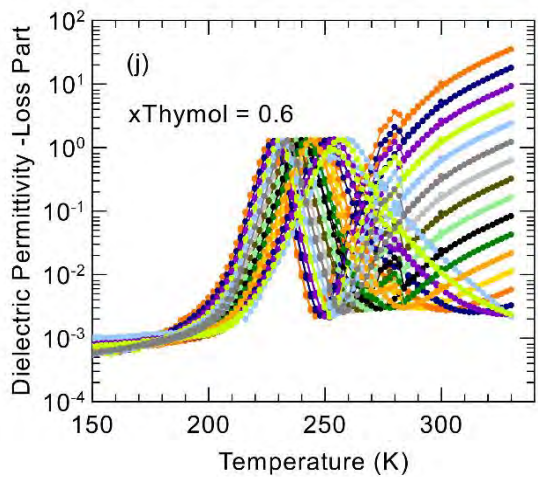
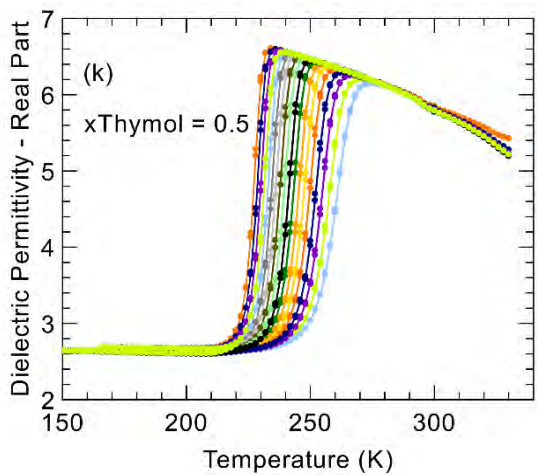
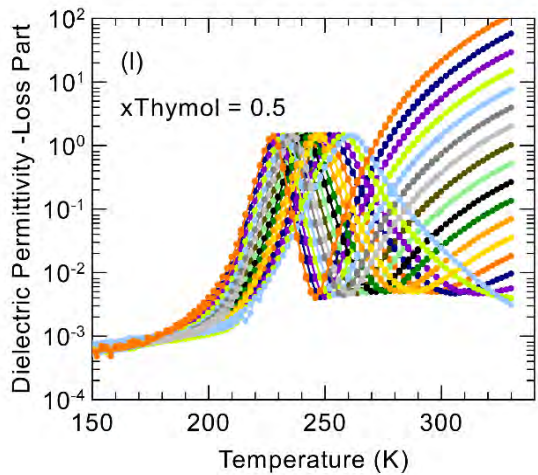

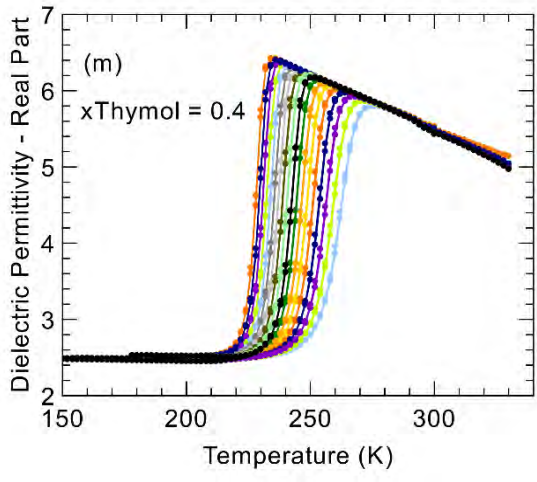
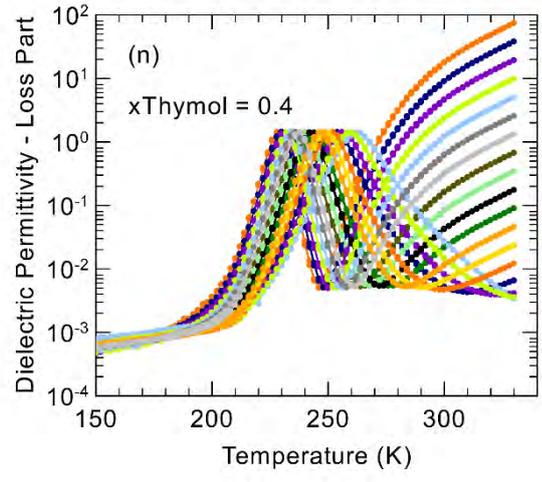
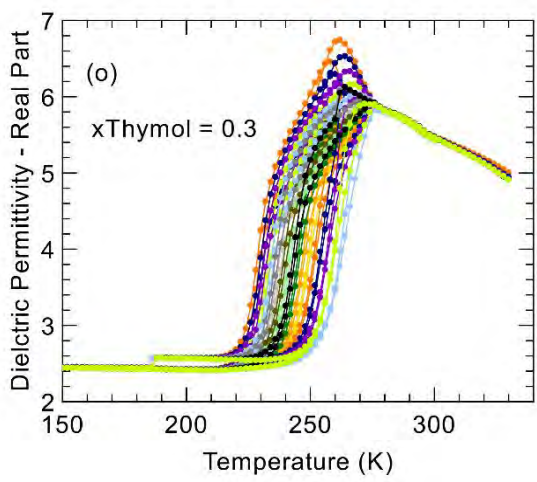
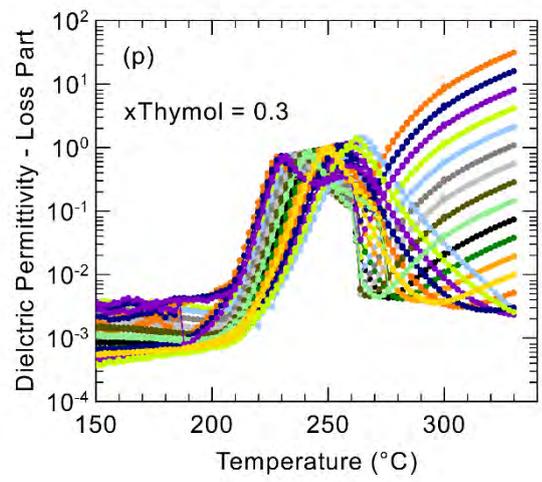
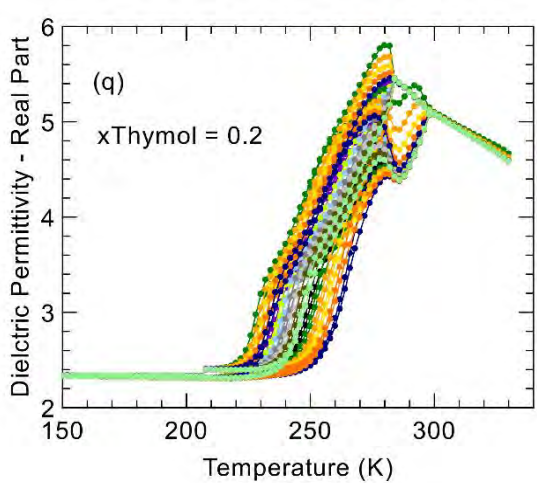
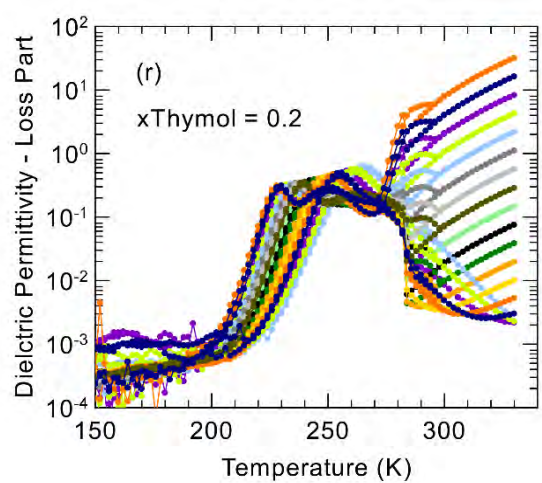

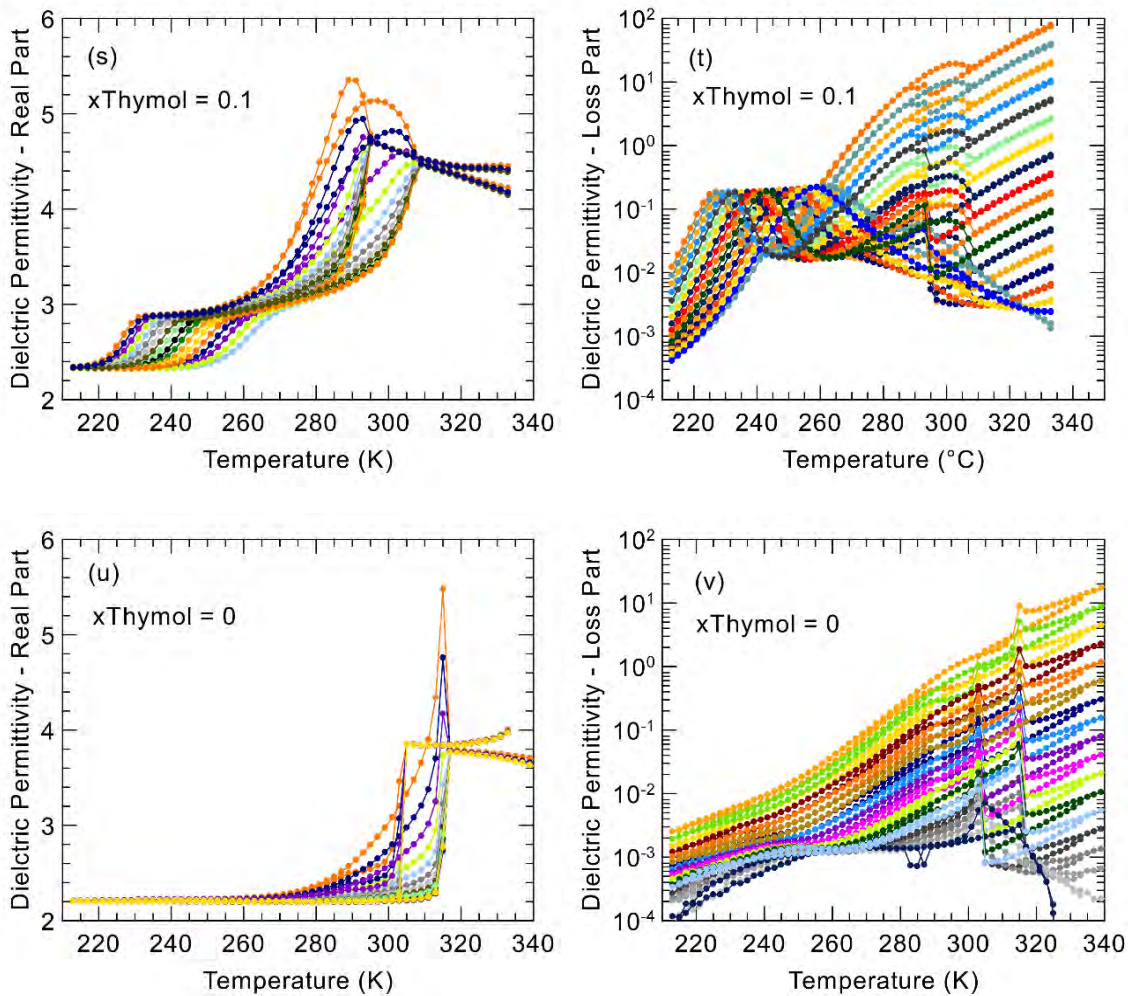

**Fig. S2.** Isochronal representation of the complex dielectric function of menthol-thymol mixtures for different values of the molar fraction of thymol going from pure thymol (upper panels) to pure menthol (lower panels). Real part (left panels) and loss part (rights panels) of the complex dielectric function are presented for a selection of frequencies going from 10 Hz to 1 MHz (cf. arrow in panel (b)). Both cooling and heating branches are presented on each panel.

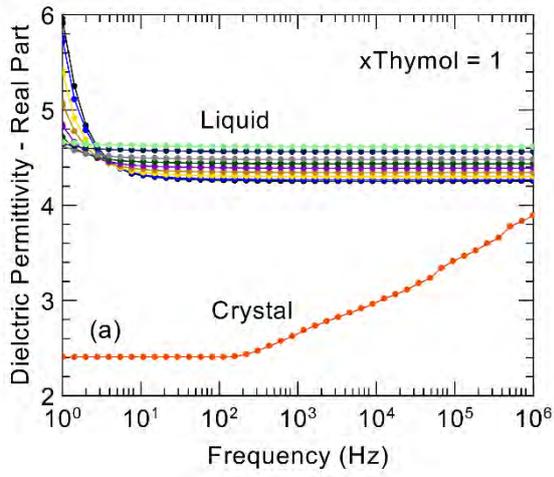
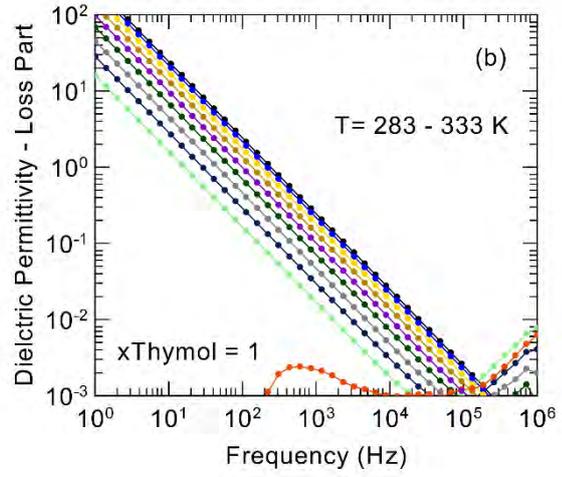
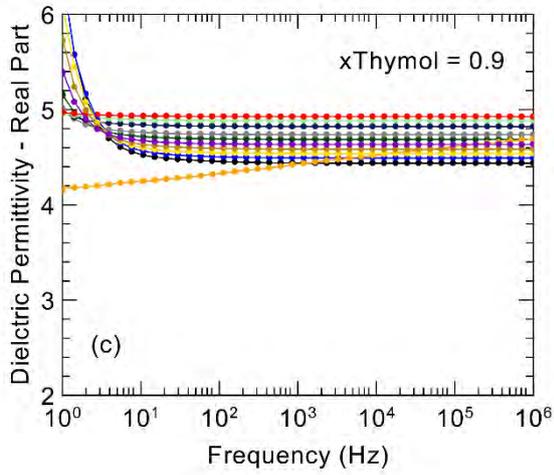
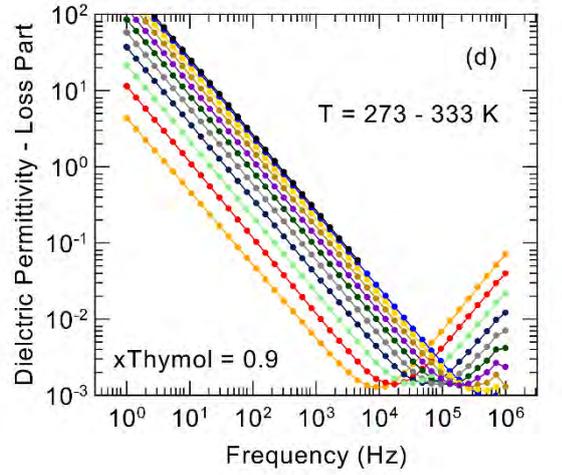
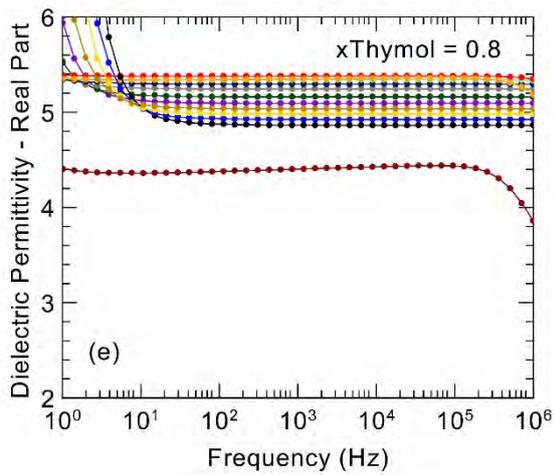
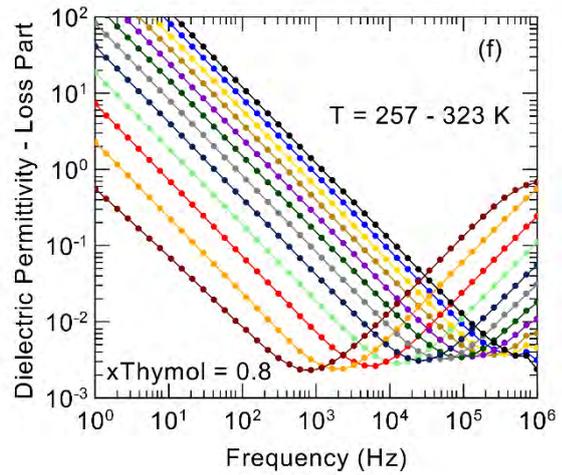

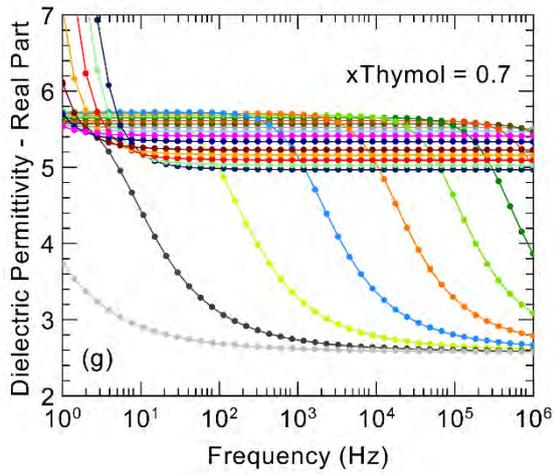
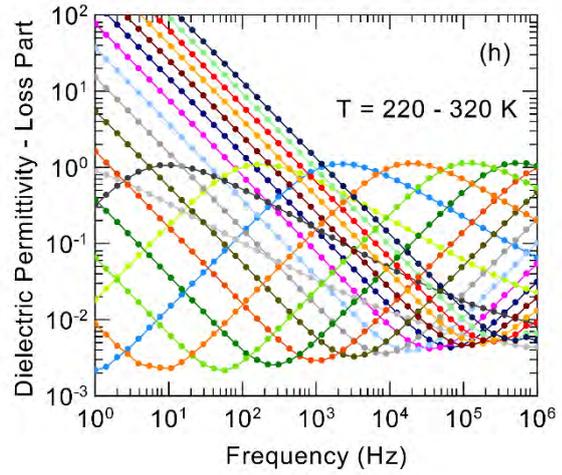
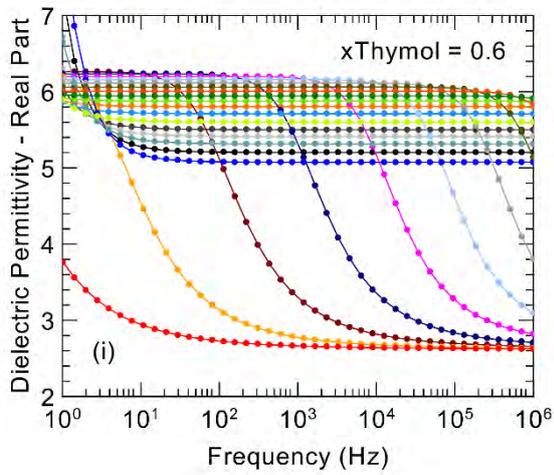
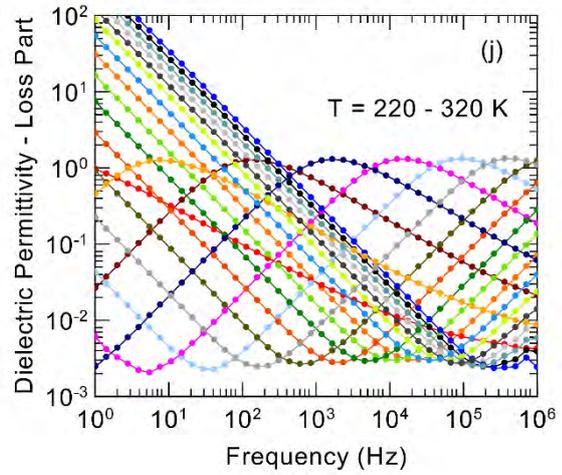
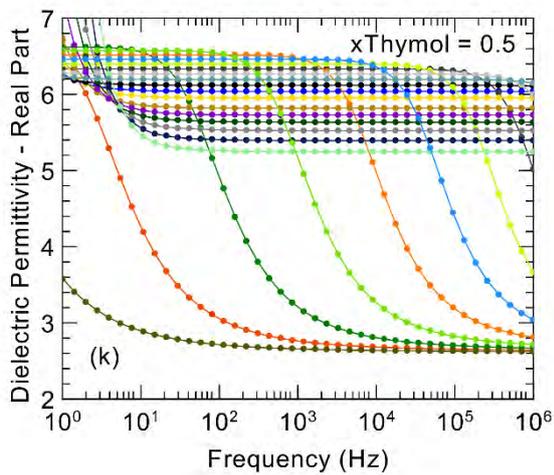
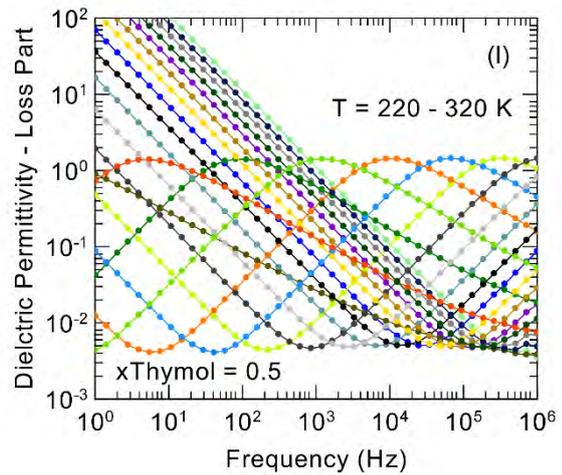

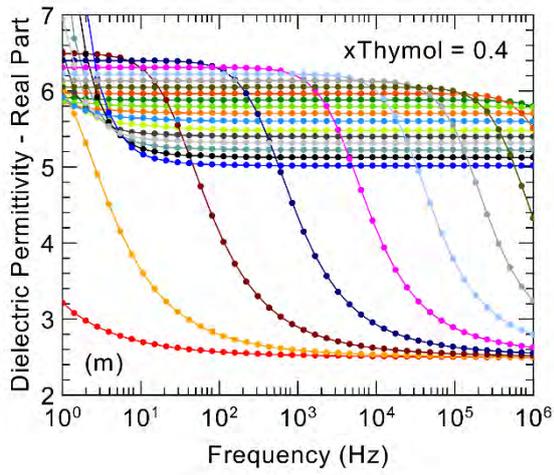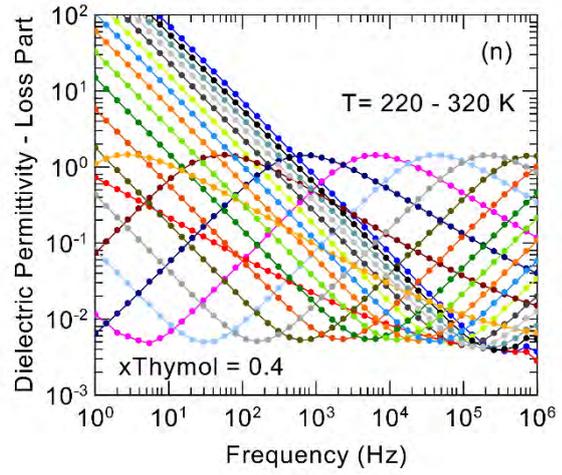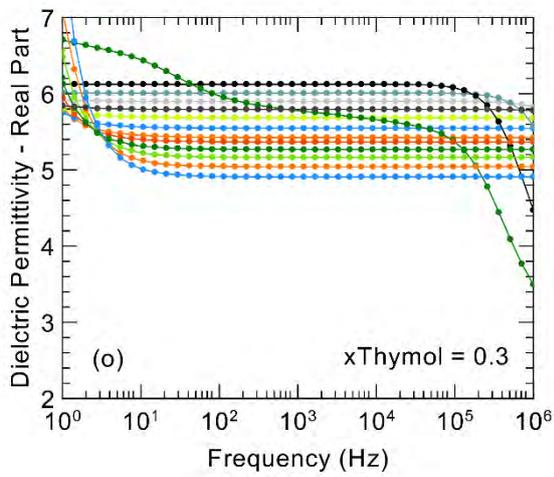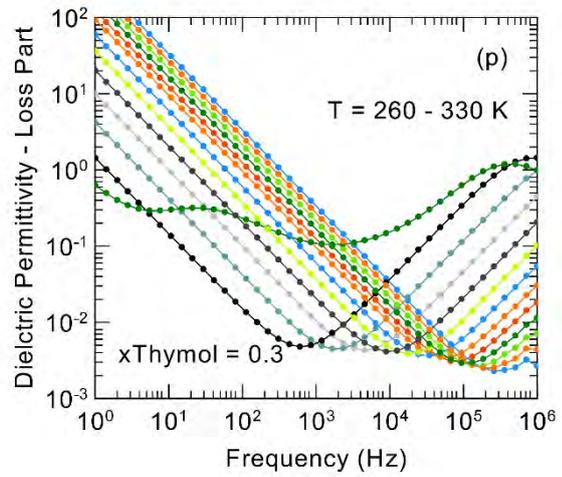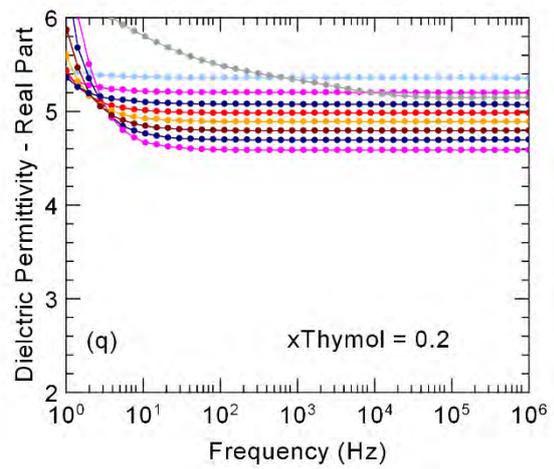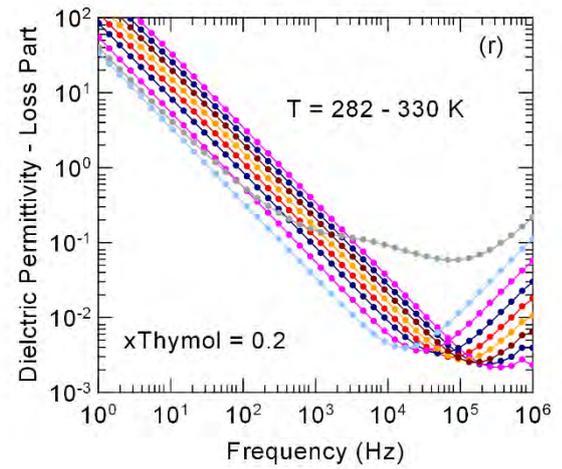

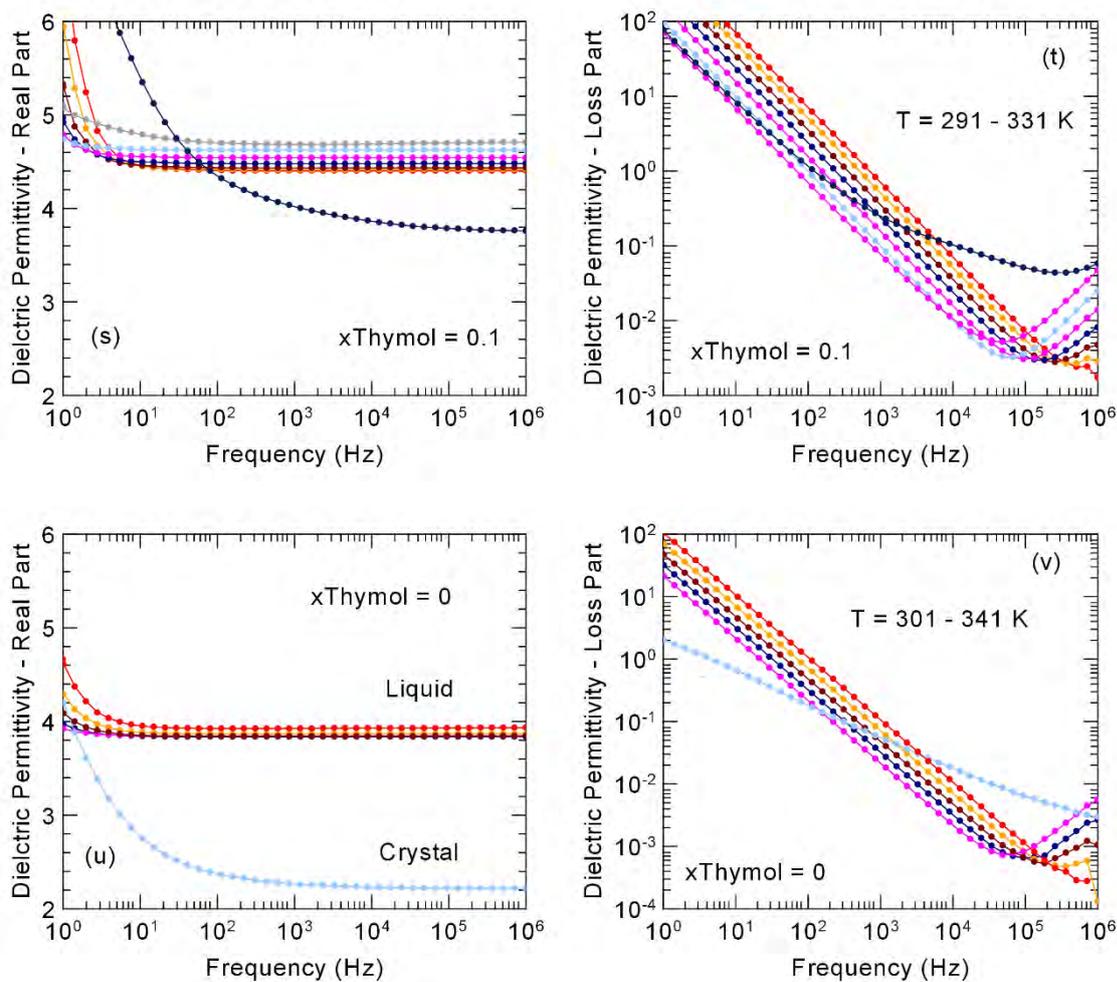

**Fig. S3.** Spectra of the complex dielectric function of menthol-thymol mixtures for different values of the molar fraction of thymol going from pure thymol (upper panels) to pure menthol (lower panels). Real part (left panels) and loss part (rights panels) of the complex dielectric function are presented for a selection of temperatures over a restricted range (indicated in right panels) corresponding to the liquid phase. A spectrum illustrating the onset of crystallization is also presented for non-glassforming compositions.